\documentclass{article}

\usepackage{arxiv}
\usepackage[utf8]{inputenc} 
\usepackage[T1]{fontenc}    
\usepackage{hyperref}       
\usepackage{url}            
\usepackage{booktabs}       
\usepackage{amsfonts}       
\usepackage{amsmath} 
\usepackage{nicefrac}       
\usepackage{microtype}      
\usepackage{cleveref}       
\usepackage{lipsum}         
\usepackage{graphicx}
\usepackage{natbib}
\usepackage{doi}
\usepackage{wrapfig,lipsum}
\usepackage{longtable}

\title{The impact of deep learning aid on the workload and interpretation accuracy of radiologists on chest computed tomography: a cross-over reader study.}

\usepackage{authblk}

\author[1]{Anvar Kurmukov}
\author[1]{Valeria Chernina}
\author[1]{Regina Gareeva}
\author[1]{Maria Dugova}
\author[1]{Ekaterina Petrash}
\author[1]{Olga Aleshina}
\author[1]{Maxim Pisov}
\author[1]{Boris Shirokikh}
\author[1]{Valentin Samokhin}
\author[1]{Vladislav Proskurov}
\author[1]{Stanislav Shimovolos}
\author[1]{Maria Basova}
\author[1]{Mikhail Goncahrov}
\author[1]{Eugenia Soboleva}
\author[1]{Maria Donskova}
\author[1]{Farukh Yaushev}
\author[1]{Alexey Shevtsov}
\author[1]{Alexey Zakharov}
\author[1]{Talgat Saparov}
\author[1]{Victor Gombolevskiy}
\author[1]{Mikhail Belyaev}

\affil[1]{AUMI.AI}

\renewcommand{\shorttitle}{Workload of radiologist}

\begin{document}
\maketitle

\begin{abstract}
Interpretation of chest computed tomography (CT) is time-consuming. Previous studies have measured the time-saving effect of using a deep-learning-based aid (DLA) for CT interpretation. We evaluated the joint impact of a multi-pathology DLA on the time and accuracy of radiologists' reading. 

40 radiologists were randomly split into three experimental arms: control (10 radiologists), unaware of DLA capabilities who interpret studies without assistance; informed group (10 radiologists), who underwent a briefing about which pathologies DLA detects, but performed readings without DLA; and the experimental group (20 radiologists), who interpreted half studies with DLA, and half without. Every arm used the same 200 CT studies retrospectively collected from BIMCV-COVID19 dataset; each radiologist provided readings for 20 CT studies. We compared participants’ interpretation time, and accuracy of their diagnostic report in terms of sensitivity and specificity with respect to 12 pathological findings.

Mean reading time per study was $15.6$ minutes  [SD $8.5$] in the control arm, $13.2$ minutes [SD $8.7$] in the informed arm, $14.4$ [SD $10.3$]  in the experimental arm without DLA, and $11.4$ minutes [SD $7.8$] in the experimental arm with DLA. Mean sensitivity and specificity were $41.5$ [SD $30.4$], $86.8$ [SD $28.3$]  in the control arm; $53.5$ [SD $22.7$], $92.3$ [SD $9.4$] in the informed non-assisted arm; $63.2$ [SD $16.4$], $92.3$ [SD $8.2$] in the experimental arm without DLA; and $91.6$ [SD $7.2$], $89.9$ [SD $6.0$] in the experimental arm with DLA. DLA speed up interpretation time per study by $2.9$ minutes ($\text{CI}_{95} [1.7, 4.3], \text{p}<0.0005$), increased sensitivity by $28.4$ ($\text{CI}_{95} [23.4, 33.4], \text{p}<0.0005$), and decreased specificity by $2.4$ ($\text{CI}_{95} [0.6, 4.3], \text{p}=0.13$).

Of 20 radiologists in the experimental arm, 16 have improved reading time and sensitivity, two improved their time with a marginal drop in sensitivity, and two participants improved sensitivity with increased time. Overall, DLA introduction decreased reading time by $20.6\%$.

\end{abstract}


\section{Introduction}

According to the Royal College of Radiology’s Workforce Census report\cite{RCR2024}, the UK National Healthcare System faces a 29\% shortfall in clinical radiologists, projected to increase to 40\% by 2027. Simultaneously, the Association of American Medical Colleges\cite{AAMC2021} anticipates a shortage between 10300 and 35600 of “Other specialities”, including radiologists, by 2034 in the US. The Radiological Society of North America also reports\cite{RSNA2022} a global shortage of radiologists. Concurrently, the amount of performed Computed Tomography studies rises, with a 7\% annual increase in the UK; a similar trend is observed in the US, where it nearly doubled from 2010 to 2020\cite{richards2022diagnostics,winder2021overdoing}. This combination of a dwindling workforce and escalating workload underscores an urgent need for enhanced efficiency.

Recent studies have showcased the potential of modern deep learning in detecting and delineating a variety of pathologies on computed tomography, magnetic resonance imaging and x-ray studies\cite{ueda2023artificial,rao2021utility,calli2021deep,shirokikh2022systematic}. While these systems match human performance metric-wise, their real-world clinical impact remains under-investigated\cite{liu2019comparison,gorenstein2023ai}. A recent review by Liu et al., highlights that although deep learning algorithms often perform comparably to healthcare professionals, few studies compare their performance on external datasets, with direct comparison with humans\cite{liu2019comparison}.

Moreover, while many studies assess the accuracy of deep-learning-based systems with application to medical imaging diagnostics, only a few authors compare the efficiency of radiologists, with and without the deep-learning-based aid (DLA) in terms of working time. Existing results suggest a 10-31\% time decrease for chest radiography, 30\% time decrease for hand radiographs.\cite{bennani2023using,ahn2022association,eng2021artificial}

Research of DLA influence on workload with respect to CT is limited. One study analysed the effect of multi-task DLA on chest CT interpretation times\cite{yacoub2022impact}. The average effect for three participant radiologists was reported to be 22.1\% time reduction. However, this study did not assess the impact on examination accuracy, and time recordings were performed manually by the participants. Another study\cite{abadia2022diagnostic} assessed a non-inferiority of AI system compared to radiology reports with 8.4\% increase in sensitivity with respect to lung nodule detection of patients with complex lung diseases. In addition, the authors demonstrated a 78\% reduced CT evaluation time (from 2:44 minutes to 35.7 seconds on average). However, the latter result was demonstrated on 20 random cases (out of 143), all of which were positive for lung cancer, based on times recorded from a single radiologist after a one-month washout.

In present work, we tested how a multi-pathology deep-learning-based aid affects radiologists’ workload and accuracy. Our study design allows for paired comparisons on the image level (the same CT study was annotated with and without DLA) and on the radiologist level (the same radiologists assessed studies with and without DLA). We demonstrated the overall positive effect of DLA examination on both workload (reduced time) and detection accuracy (increased sensitivity, with preserved level of specificity). 

\section{Methods}

We evaluated a deep-learning-based software designed to assist in diagnosing 12 chest and abdomen pathologies on CT scans, including lung nodules, features of viral pneumonia, emphysema and pleural effusion, lymphadenopathy in the intrathoracic lymph nodes, aorta and pulmonary trunk enlargement, coronary calcium, adrenal lesions, ribs and vertebrae fractures, and vertebrae mineral density. We performed paired comparisons in two distinct ways: first, by enabling the same radiologists to evaluate different CT studies with and without DLA assistance, and second, by allowing various radiologists to assess the same CT studies, both with and without the DLA. We measured the joint effect on workload and performance and analysed inter-/intra-readers and inter-group variability.

\subsection{Software}

EfficientReadCT (version 1.0.0, AUMI.AI) was used for the purposes of this study. It is a commercially available deep-learning-based computer aid diagnosis system developed for detection of various pathologies on CT studies. It automates morphological measurements and provides diagnostic suggestions for 12 pathologies in a structured form according to the international guidelines\cite{
macmahon2017guidelines,francone2020chest,occhipinti2019spirometric,karkhanis2012pleural,munden2018managing,ESC,glazer2020management,he2019ideal,genant1993vertebral,ISCD2023}, see Supplementary materials Table 2.

\subsection{Study design}

\begin{figure}
\begin{center}
\includegraphics[width=1\linewidth]{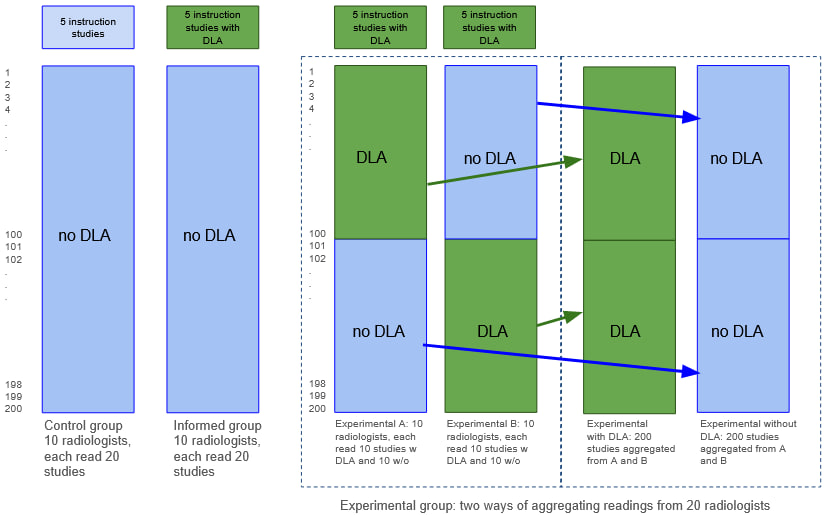}
\end{center}
\caption{Study design.}
\label{fig:study-design}
\end{figure}
The primary aim of this study was to estimate the effect of DLA interpretation on CT reading times. The secondary goal was to estimate the effect on examination accuracy. Radiologists were not informed that their reading time was being recorded. They were informed that the diagnostic accuracy of their readings would be measured. They were paid regardless of their performance.

Radiologists assessed the studies from their personal computers using the RadiAnt DICOM viewer (v.2023.1). Each study was accompanied by patient age, sex and anamnesis and diagnostics task e.g. “49 y.o. woman with cough and expectoration”. For each study, the participants were asked to fill up a web form pre-filled either with a standard report describing the organs and systems included in the scan, without pathological findings, or report with DLA results, see examples in Supplementary Table 1. After interpreting a study, the participant submitted the report and was not able to edit it later. Time spent on the interpretation of a single study was automatically recorded as the time between pressing the “Start new study” and “Submit a report” buttons in the webform. Participants were asked not to take breaks during the interpretation of a single study but could make any number of breaks between the studies. Participants were instructed to finish all their studies in a single working day.  

We enrolled 40 radiologists and divided them into three groups, consisting of 10, 10 and 20 radiologists, see Figure 1 and group descriptions below. Each participant annotated 20 unique CT studies. In total, we used 200 unique CT studies for the experiment; each study was annotated independently by four radiologists.

CT studies were selected from publicly available data collection approved for research purposes.27 Studies interpretation was performed retrospectively without affecting patient treatment plans.

\subsubsection{Groups without DLA}
The first 10 radiologists, mean working experience 9·8 years [SD 2·8], were a control group who annotated the experimental data without DLA. The purpose of this group was three-fold. First, to estimate the baseline reading time and accuracy. Second, to compare the performance of the participants with the accuracy of the original radiological reports. This established the effect of non-clinical/non-prospective experimental design. Third, to compare with radiologists in the informed group. 

The second group of 10 radiologists, mean working experience 8.3 years [SD 3.1], i.e. “informed group” also annotated experimental data without the DLA. However, before the experiment, they received five CT studies processed by the DLA which covered all pathologies known to it. The purpose of these five studies was to familiarise participants with the DLA capabilities which include pathologies rarely mentioned in the readings during routine examination, such as loss of vertebrae mineral density, adrenal incidentalomas, and others. We hypothesised that radiologists who are aware that the DLA detects such pathologies would also pay more attention to them, even without the DLA.

\subsubsection{The experimental group}
The experimental group consisted of 20 radiologists, mean working experience 9.3 years [SD 3.4], each interpreted studies with and without the DLA. Before the experiment, all participants from this group received five CT studies with deep learning-generated overlay. During the experiment, each participant annotated half of their studies with the DLA, and half - without. Studies with and without the DLA were read in random order to mitigate the potential effect of fatigue to the end of the experiment session\cite{taylor2019fatigue}. Studies were distributed in such a way, that each of 200 studies were once annotated with the DLA, and once - without.

\begin{figure}[!tbp]
  \centering
  \begin{minipage}[b]{0.49\textwidth}
    \includegraphics[width=\textwidth]{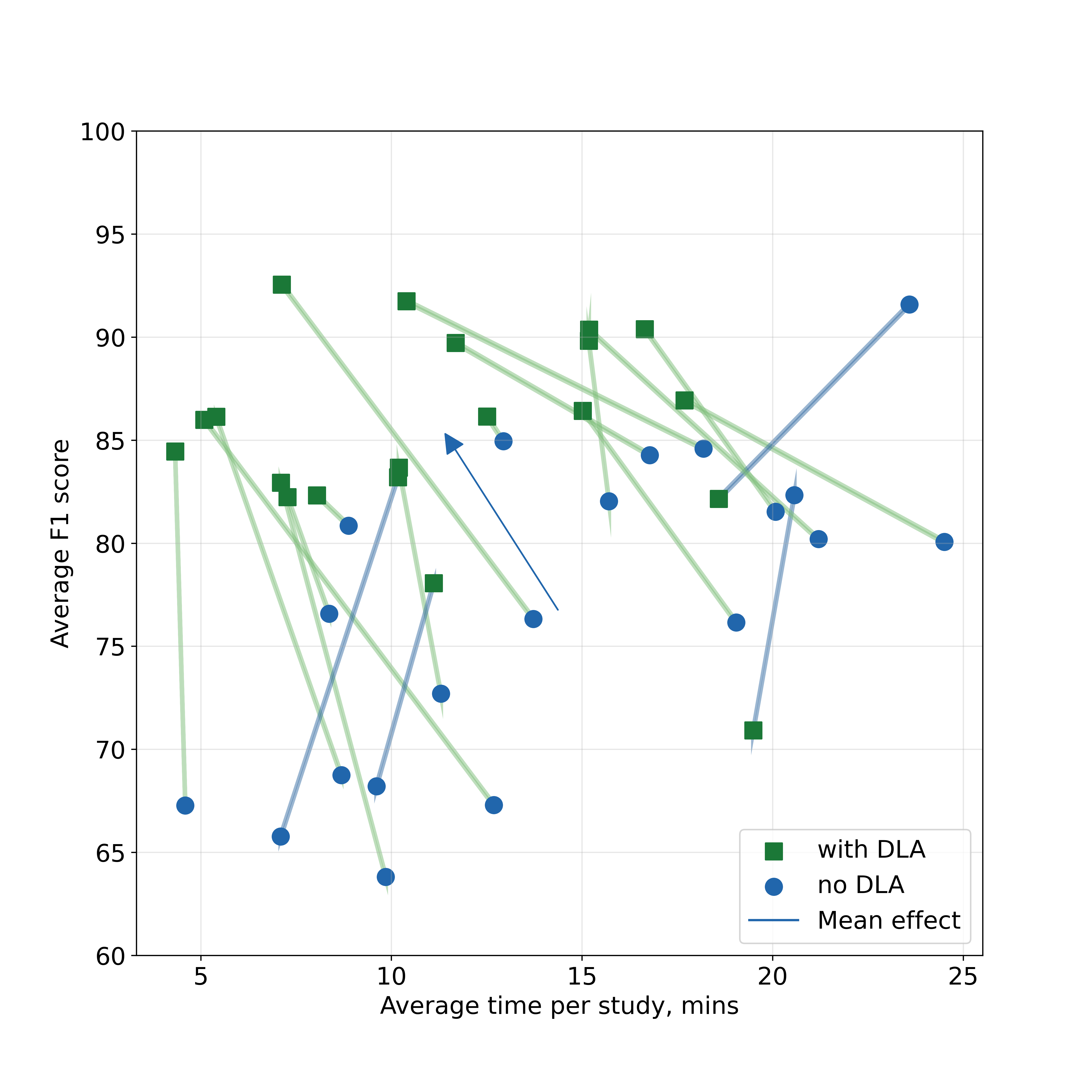}
    \caption{Time vs F1 score in experimental group by participants. Each line represents one participant from the experimental arm.}
  \end{minipage}
  \hfill
  \begin{minipage}[b]{0.49\textwidth}
    \includegraphics[width=\textwidth]{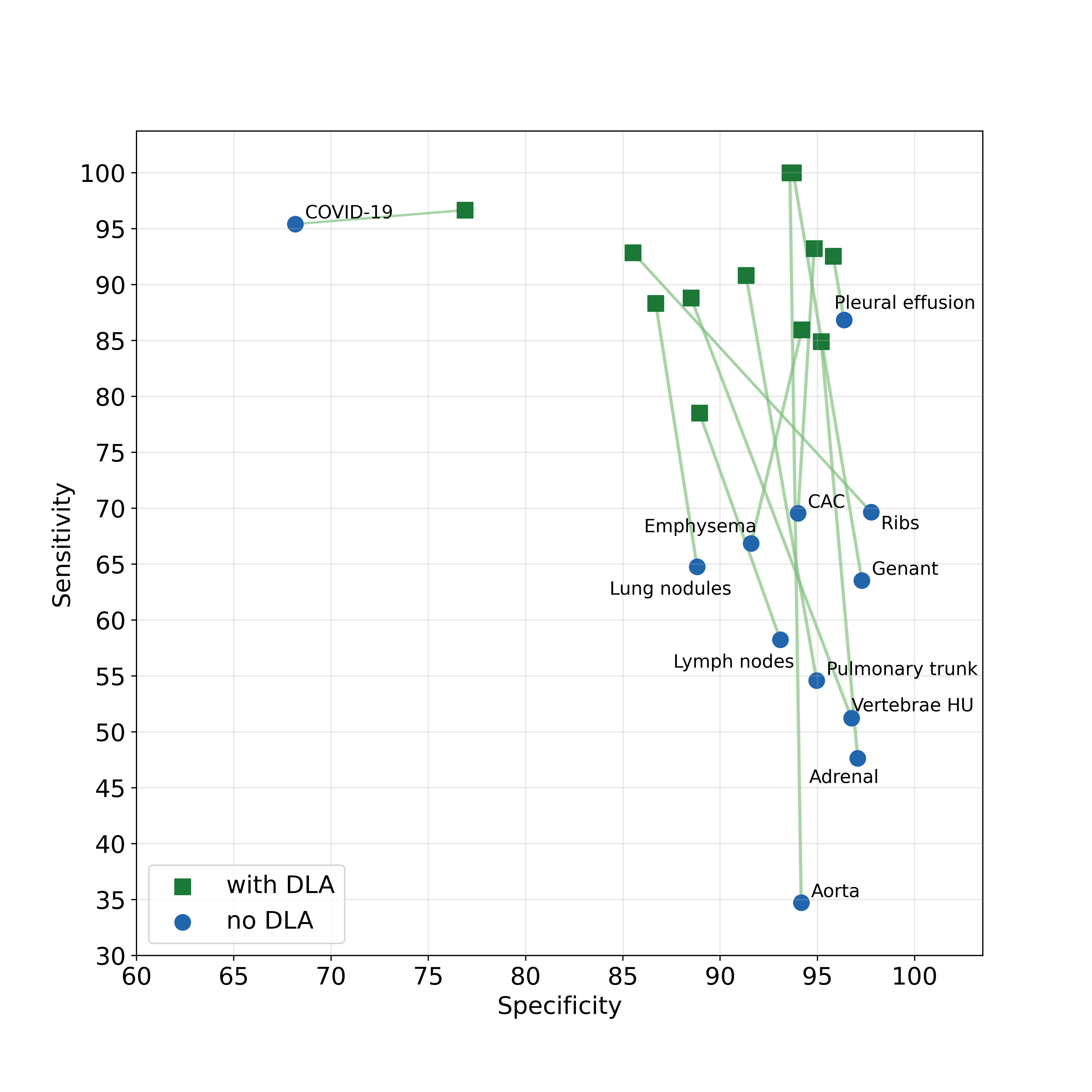}
    \caption{Sensitivity and specificity by pathologies between participants with DLA and without in the experimental arm.}
  \end{minipage}
\end{figure}

The experimental group could be seen in two ways, see Figure 1. First, as two independent groups: experimental A and experimental B each consists of 10 participants. Each radiologist in either group interpreted 20 unique studies (10 with the DLA and 10 without). For each radiologist this allowed us to compare the performance and reading time on 10 studies without the DLA versus 10 studies with DLA (a crossover design). 

Second, if a study was analysed with the DLA in group A, it was analysed without the DLA in group B. Thus, the experimental group could be seen as 200 studies analysed once without the DLA (by 20 radiologists, each provided 10 readings) and once with the DLA (by the same 20 radiologists, each provided 10 readings). We used these two groups to derive the effects of the DLA introduction.

\subsection{Validation data}
We used CT studies from the BIMCV COVID-19+ collection\cite{vaya2020bimcv}.  We selected 205 CT studies without contrast enhancement with chest region (Th1 to Th12 vertebrae present). Of 205 subjects, five studies were used to familiarise participants with the experiment interfaces, and the DLA capabilities in case of informed and experimental groups. The remaining 200 studies were used for the experiment, of which 97 (48.5\%) were male, mean age 67.7 years [SD 14.5], and 103 (51.5\%) were female, mean age 62.7 [SD 16.9]. We manually extracted BIMCV annotations from medical reports associated with the CT studies, see details in the Supplementary materials “Text annotation”. 

\subsubsection{Ground truth development}
To obtain the ground truth labels, we compared six sources of labels: extracted from the original medical reports, from radiologists from each experimental arm, and generated by the DLA  system. If all six sources agreed on a specific diagnosis, we accepted it as the ground truth. Otherwise, if any of the six sources disagreed, the study was reviewed by an experienced radiologist for this particular pathology without DLA. 

The annotation of severity of pathologies (e.g. aorta enlargement > 50 mm, Agatston’s score > 300) was performed in a second round of annotation with access to DLA. Pleural effusion was excluded from the severity analysis, because it often requires the analysis of drained fluids, which was not possible in the retrospective setup.

\begin{table}[ht]
\centering
\caption{International guidelines on selected pathologies diagnostics. Criteria for lymphadenopathy, aorta enlargement, vertebrae fractures and COVID-19 were simplified.}
\resizebox{0.95\textwidth}{!}{
\renewcommand{\arraystretch}{1.5}
\begin{tabular}{lllll}
\toprule
Pathology & Location & low & moderate & severe \\
\midrule
lung nodules \cite{macmahon2017guidelines} & Chest & <6 mm (100 mm\textsuperscript{3}) & 6-8 mm (100-250 mm\textsuperscript{3}) & >8 mm (>250 mm\textsuperscript{3}) \\
COVID-19 \cite{francone2020chest} & Chest & 1-25\% lobar involvement & 25-50\% lobar involvement & >50\% lobar involvement\\
emphysema \cite{occhipinti2019spirometric} & Chest & \begin{tabular}[c]{@{}l@{}}isolated bullas < 10mm,\\ $\text{LAA}_{-950\text{insp}}$ 1-5\%\end{tabular} & $\text{LAA}_{-950\text{insp}}$  6-14\% & $\text{LAA}_{-950\text{insp}}$ $\geq$ 14\% \\
lymphadenopathy \cite{munden2018managing} & Chest & short axis < 10mm & short axis 11-15 mm & short axis >15 mm \\
ascending aorta \cite{ESC} & Chest & - & diameter 40-49 mm & diameter  >50 mm \\
descending aorta \cite{ESC} & Chest & - & diameter 31-39 mm & diameter  >40 mm \\
descending aorta \cite{ESC} & Abdomen & - & diameter  26-30 mm & short axis >30 mm \\
pulmonary artery \cite{munden2018managing} & Chest & diameter  29-30 mm & diameter  31-33 mm & diameter  31-33 mm \\
coronary calcium \cite{munden2018managing} & Chest & Agatston score 1-100 & Agatston score 101-300 & Agatston score >300 \\
adrenal lesions \cite{glazer2020management} & Abdomen & \begin{tabular}[c]{@{}l@{}}short axis < 10mm,\\ density <10HU\end{tabular} & \begin{tabular}[c]{@{}l@{}}short axis 11-39mm,\\ density  >10HU\end{tabular} & \begin{tabular}[c]{@{}l@{}}short axis >40mm,\\ density  >10HU\end{tabular} \\
ribs fractures \cite{he2019ideal} & Bones & consolidated fracture & nondisplaced fracture & displaced fracture \\
vertebral fractures\cite{genant1993vertebral} & Bones & height reduction 20-25\% & height reduction 26-40\% & height reduction >40\% \\
vertebral density \cite{ISCD2023} & Bones & 100-150 HU & <100 HU & - \\
\bottomrule
\end{tabular}}

\label{tab:guidelines}
\end{table}

Dataset statistics on the number of pathologies with different severity are available in the Supplementary materials Tables 3-4.

\subsection{Statistical analysis}
Statistical analysis were performed in Python, Wilcoxon signed-rank test was
used for paired comparisons. Two-sided 95\% confidence intervals were assessed via bootstrapping (over studies) with 1000 repetitions. We set the alpha significance level at $0.03$.

\section{Results}
\subsection{BIMCV vs Control vs Informed}
First, we compare the control group with the results derived from the original medical reports. For most of the analysed pathologies we observe no statistically significant differences in sensitivity and specificity, except for emphysema, coronary calcium and lymph nodes, for which radiologists in the control group demonstrate higher sensitivity; lung nodules, for which original reports have higher sensitivity; and infiltration and consolidation of lung tissue during COVID and lung’s nodules, for which original reports have higher specificity. Averaged over 12 pathologies, sensitivity is higher in the control group by 10.2 points (CI $[5.5, 14.3]$,  $p=0.026$), there is no statistically significant difference in specificity (CI $[0.1, 2.6]$, $p=0.23$), see Table 1. Tables with pathology-wise classification metrics  for all arms are provided in the Supplementary materials Table 5.

Second, we compare the control and informed groups. On average sensitivity is higher in the informed group by $11.9$ points (CI $[7.1, 17.1]$, $p=0.016$), the differences in specificity are not statistically significant (CI $[2.0, 5.4]$, $p=0.16$). This result supports our initial hypothesis that familiarity with DLA pathologies increases participants' awareness of them.

Finally, we report that reading times were 2.4 fewer minutes (CI $[0.8, 4.0]$, $p=0.0007$)  in the informed group (on average per study). We attribute this result to individual differences among radiologists in interpretation time, rather than to the familiarity with DLA, see next section.

\subsection{Experimental group}
\subsubsection{Inter-reader variability}
First, we compare the Experimental groups A and B in terms of time and performance to measure inter-radiologists’ differences, while preserving the reading conditions, see Table 1. Both groups analysed the same 200 CT studies, half with DLA and half without. There were no statistically significant differences between the groups in terms of sensitivity (CI $[-9.5, 4.1]$, $p=0.57$), and specificity (CI $[-0.7, 2.8]$, $p=0.34$). However, participants in Experimental group A spent on average 4.06 more minutes (CI $[2.6, 5.4]$, $p<0.0001$) on reading a single study. These results demonstrate that there is low inter-group variability in performance but high variability in time, meaning that the effect on workload could only be measured in a cross-over fashion, i.e., comparing reading time of radiologists with themselves with and without DLA.

\begin{table}[ht]
\centering
\caption{Time and performance metrics of radiologists in different experimental groups. Mean (std) are averaged over 12 pathologies. Time mean (std) averaged over studies.}
\begin{tabular}{lccccc}
\toprule
Group/Metrics & Sensitivity & Precision & Specificity & F1 & Time, mins \\
\midrule
BIMCV annotation & 31.3 (30.7) & 77.8 (20.0) & 96.5 (5.5) & 37.4 (27.5) & n/a \\
Control & 41.5 (30.5) & 76.0 (22.6) & 95.2 (7.3) & 51.7 (25.5) & 15.6 (8.5) \\
Informed & 53.5 (22.7) & 73.2 (16.3) & 91.5 (9.4) & 57.8 (16.8) & 13.2 (8.7) \\
Experimental A & 76.1 (11.5) & 77.0 (15.4) & 91.6 (7.8) & 75.5 (10.7) & 10.9 (7.5) \\
Experimental B & 78.7 (9.9) & 75.8 (12.2) & 90.6 (5.9) & 76.3 (7.3) & 14.9 (10.4) \\
Experimental without DLA & 63.2 (16.4) & 77.1 (13.7) & 92.3 (8.2) & 67.8 (11.1) & 14.4 (10.3) \\
Experimental with DLA & 91.6 (7.2) & 76.4 (14.3) & 89.9 (6.1) & 82.5 (9.5) & 11.4 (7.8) \\
\midrule
Automated (no radiologist) & 89.9 (8.3) & 73.0 (14.4) & 87.3 (9.2) & 79.5 (9.1) & n/a \\
\bottomrule

\end{tabular}

\label{tab:metrics}
\end{table}

\subsubsection{Time and performance}
Figure 2 demonstrates the joint effect on performance and  workload from DLA introduction. 16 out of 20 participants benefited in terms of both time and F1 score, two participants substantially increased their F1 score, with increased workload, two participants examined studies faster with DLA, but with lower average F1 score. Overall, while using DLA participants had higher sensitivity by 28.4 points (CI $[23.5, 33.5]$, $p=0.0005$), differences in specificity were not statistically significant (CI $[0.7, 4.3]$, $p=0.13$), and  F1 score is higher by 14.8 points (CI $[10.7, 18.7]$, $p=0.0010$).

The average time saved per study was $2.9$ minutes (CI $[1.7, 4.3]$, $p=0.0005$) per study, or 20.6\% (CI $[14.9\%, 37.7\%]$). On an individual radiologists level, DLA introduction decreased reading time for 18 out of 20 participants, see Figure 2. For each of 18 radiologists, the decrease in examination time was statistically significant with $p<0.0001$. Figure 3 demonstrates the effect on individual pathologies, the increase in sensitivity is especially apparent for pathologies determined by morphological measurements, such as the diameter of the aorta or pulmonary artery, vertebrae fractures and density. In Supplementary materials we provide radiologists-wise metrics for all groups, Table 6.

\subsubsection{Findings severity}
To analyse how radiologists’ accuracy changes depending on the severity of pathologies, we computed  stratified  sensitivity, see Table 3. We do not report Specificity as True Negatives could not be classified for severity. We also do not report Precision, because when a radiologist describes a finding which is not present (a False Positive), it is too arbitrary to decide, which severity level was implied from the text description.  

\begin{table}[ht]
\centering
\caption{Sensitivity stratified by the findings' severity. Mean (std) are averaged over 12 pathologies.}
\begin{tabular}{lccc}
\toprule
& low & moderate & severe \\
\midrule
BIMCV annotation & 24.8 (33.4) & 35.4 (37.2) & 47.2 (35.1) \\
Control & 39.2 (30.9) & 48.1 (41.4) & 60.6 (25.1) \\
Informed & 40.3 (28.1) & 64.0 (30.8) & 67.9 (27.9) \\
Experimental A & 65.4 (21.1) & 81.2 (17.3) & 89.1 (12.1) \\
Experimental B & 64.8 (23.0) & 81.5 (16.3) & 85.7 (10.6) \\
Experimental without DLA & 48.0 (29.5) & 69.2 (20.2) & 81.0 (16.1) \\

Experimental with DLA & 82.2 (15.8) & 93.5 (11.8) & 93.8 (7.3) \\
\midrule
Automated (no radiologist) & 70.7 (29.8) & 93.8 (8.4) & 93.3 (12.3) \\
\bottomrule
\end{tabular}
\label{tab:severity}
\end{table}

As expected, the sensitivity of radiologists increases when dealing with conditions of higher severity, in every arm. For findings of low severity, we observe an increase in sensitivity from 39.4 (CI $[0.0, 98.3]$) in the control group, to $48.1$ (CI $[0.0, 96.6]$) in the experimental group without DLA, and to $82.3$ (CI $[28.6, 100.0]$) in the experimental group with DLA. For moderate severity findings we observe an increase in sensitivity from $44.5$ (CI $[0.0, 100.0]$), to $66.3$ (CI $[26.7, 100.0]$) and 94.0 (CI $[57.1, 100.0]$) in the control group, experimental group without DLA and with DLA respectively. 

Finally, for findings of high severity, sensitivity increased from 60.6 (CI $[10.3, 100.0]$) in the control group, to 80.9 (CI $[40.0, 100.0]$) in the experimental group without DLA and $93.8$ (CI $[66.7, 100.0]$) in the experimental group with DLA.

Wide confidence intervals are due to the relatively low number of positive findings for each severity, see Supplementary materials for severity statistics in Table 3.

\section{Discussion}

Numerous studies have explored the accuracy of AI algorithms applied to radiography; however, only a handful have compared the efficiency of radiologists, with and without DLA assistance, in terms of working time. Eng et al. reported a 30\% decrease in workload time for skeletal age assessment from hand radiographs, alongside a 9·5\% increase in accuracy\cite{eng2021artificial}. A similar estimate of 31\% saved time was reported by Bennani and colleagues, who enlisted 12 radiologists with varying work experience to analyse five lung pathologies on chest X-rays\cite{bennani2023using}. Conversely, Ahn et al. reported a smaller effect of a 10\% decrease in workload from AI for similar tasks, albeit with a significant increase in sensitivity\cite{ahn2022association}. These studies underscore the potential of AI to enhance both efficiency and accuracy in radiography.

The impact of deep learning aid assistance on workload for CT readings remains underexplored. Abadia et al. demonstrated the non-inferiority of an AI system compared to radiology reports for detecting lung cancer in patients with complex lung diseases\cite{abadia2022diagnostic}. They also reported significantly reduced CT evaluation times. However, the latter result was demonstrated using a limited number of cases based on reading times of a single radiologist. Yacoub et al. explored the effects of a multi-pathology DLA system for thoracic and abdominal CT on radiologists’ working time, reporting a 22\% reduction in workload from three radiologists\cite{yacoub2022impact}. Main limitation of this study is that the authors did not address the potential performance trade-off associated with reduced working time.

In this work, we demonstrated that DLA introduction reduces interpretation time per study by 2.9 minutes (20.8\%), simultaneously increasing sensitivity by 28.4, and preserving the same level of specificity. Importantly, we showed that the effect of introducing DLA assistance, is of the same magnitude as the effect of changing the radiologist, which is 4.06 minutes (CI $[2.6, 5.4]$). This result underlines the importance of radiologists-wise cross-over study design. We also demonstrated that while all of selected pathologies benefited from DLA introduction in sensitivity, pathologies diagnosed based on morphological measurements benefited more.

Our study has several limitations. First, we used retrospective data. Second, findings derivation methodology focused on all pathologies present on image, this resulted in relatively low metrics of annotations extracted from original medical reports. The discordance in accuracy between findings extracted from medical reports and findings described from CT readings retrospectively is a known phenomena\cite{gatt2003chest}. Some of the findings that we included in the analysis might not have been clinically relevant for the specific patient, such as trace amounts of pleural effusion, decrease in bone density or consolidated ribs fractures. Especially because we used retrospective data from the emergency department from the COVID-19 pandemic period. To estimate the magnitude of the non-clinical retrospective design we included a control group of 10 radiologists who interpreted the scans without any aid or knowledge about DLA capabilities, and found the effect to be about 10.2 points of sensitivity. 
Third, we did not perform a power analysis prior to study start. However, the number of participants and studies in our experiment matches or outnumbers similar works on CT and chest radiography\cite{bennani2023using,ahn2022association,yacoub2022impact,abadia2022diagnostic}.
Finally, an interesting direction for future work would be to investigate the collateral effect of DLA systems on the accuracy of radiologists' diagnoses of pathologies not included in the DLA scope. This could provide further insights into the potential and limitations of DLA systems in radiology. 

\subsection{Conclusion}

Current development of computer vision deep-learning-based AI systems for pathology detection and morphology annotation is approaching the best of human experts and outperforms an “average” radiologist who works without computer aid assistance. Radiologists augmented with DLA systems spend less time on CT examination and are more accurate with regard to pathologies highlighted by DLA. Results from independent research groups suggest a similar estimate of a 20\% workload reduction effect from multi-pathology DLA introduced into clinical practice.

In this study we demonstrated that the use of deep-learning-aid for CT interpretation decreases total time spent on CT interpretation and increases sensitivity for 12 pathologies included in the DLA scope: lung nodules, features of viral pneumonia, emphysema and pleural effusion, lymphadenopathy, aorta and pulmonary trunk enlargement, coronary calcium, adrenal lesions, ribs and vertebrae fractures, and vertebrae mineral density. These results suggest that the integration of deep-learning-aid in radiology practice holds great promise for improving efficiency and diagnostic accuracy, ultimately benefiting patient care.

\subsection{Contributors}

Study design: Mikhail Beliaev, Valeria Chernina, Maria Dugova, Ekaterina Petrash, Anvar Kurmukov. Conceptualisation: Mikhail Beliaev, Regina Gareeva, Victor Gombolevskiy. Data management: Maxim Pisov, Vladislav Proskurov, Anvar Kurmukov, Maria Basova, Maria Dugova, Ekaterina Petrash, Valeria Chernina, Olga Aleshina. Software development: Maxim Pisov, Boris Shirokikh, Valentin Samokhin, Stanislav Shimovolos, Mikhail Goncharov, Eugenia Soboleva, Maria Donskova, Farukh Yaushev, Alexey Shevtsov, Alexey Zakharov, Talgat Saparov. Statistical analysis: Anvar Kurmukov, Maria Basova. Writing: Anvar Kurmukov, Mikhail Beliaev, Victor Gombolevskiy.

\subsection{Data Sharing}
Validation data are a subset of a public dataset BIMCV COVID-19+, models’ training data and pretrained models’ weights will not be made publicly available due to intellectual property-related constraints. Studies unique identifiers, records collected during the experiment (per-radiologist examination time, examination labels, ground truth annotations), are publicly available at \url{https://zenodo.org/doi/10.5281/zenodo.10965415}. 

\bibliographystyle{unsrtnat}
\bibliography{references}
\section{Supplementary}

\subsection{Experiment report template}
Participants in each arm filled up a structured report in an editable web form during CT reading. In the case of Control and Informed arms, the report for each CT study was pre-filled with the standard “no pathologies present” template, see Table \ref{tab:experiment_form_templates}. In the experimental arm, while working with DLA, it was prefilled with DLA findings, see example in Table \ref{tab:experiment_form_templates}. 

\begin{table}[ht]
\centering
\caption{Total number of findings by pathology in different groups.}
\resizebox{0.99\textwidth}{!}{
\begin{tabular}{lccccccccccccc}
\toprule
Group/pathology & \begin{tabular}[c]{@{}c@{}}Lymph\\ nodes\end{tabular} & \begin{tabular}[c]{@{}c@{}}Coronary\\ calcium\end{tabular} & \begin{tabular}[c]{@{}c@{}}Pulmonary\\ trunk\end{tabular} & COVID & Ribs & \begin{tabular}[c]{@{}c@{}}Pleural\\ effusion\end{tabular} & Adrenal & Genant & \begin{tabular}[c]{@{}c@{}}Lung\\ nodules\end{tabular} & Emphysema & Osteoporosis & Aorta & Total \\
\midrule
Ground Truth & 45 & 120 & 66 & 73 & 29 & 51 & 40 & 27 & 27 & 67 & 109 & 42 & 696 \\
Control (no AI) & 27 & 73 & 17 & 100 & 8 & 45 & 6 & 9 & 40 & 48 & 0 & 9 & 382 \\
With AI-aid training & 41 & 71 & 45 & 113 & 12 & 47 & 19 & 18 & 39 & 58 & 14 & 31 & 508 \\
Experimental without AI & 36 & 91 & 43 & 111 & 22 & 48 & 24 & 23 & 38 & 59 & 54 & 24 & 573 \\
Experimental with AI & 53 & 116 & 72 & 103 & 54 & 53 & 40 & 38 & 46 & 65 & 118 & 52 & 810 \\
BIMCV annotation & 24 & 24 & 12 & 93 & 3 & 43 & 7 & 4 & 27 & 23 & 1 & 8 & 269 \\
AI & 57 & 103 & 67 & 100 & 61 & 65 & 44 & 37 & 48 & 66 & 118 & 53 & 819 \\
\bottomrule
\end{tabular}}
\label{tab:total_findings_by_pathology}
\end{table}
\begin{table}[ht]
\centering
\caption{Number of ground truth findings stratified by different severity levels.}
\resizebox{0.99\textwidth}{!}{\begin{tabular}{lccccccccccccc}
\toprule
Severity & \begin{tabular}[c]{@{}c@{}}Lymph\\ nodes\end{tabular} & \begin{tabular}[c]{@{}c@{}}Coronary\\ calcium\end{tabular} & \begin{tabular}[c]{@{}c@{}}Pulmonary\\ trunk\end{tabular} & COVID & Ribs & \begin{tabular}[c]{@{}c@{}}Pleural\\ effusion\end{tabular} & Adrenal & Genant & \begin{tabular}[c]{@{}c@{}}Lung\\ nodules\end{tabular} & Emphysema & Osteoporosis & Aorta & Total \\
\midrule
Low & 4 & 42 & 14 & 52 & 29 & 31 & 0 & 0 & 5 & 52 & 51 & 0 & 280 \\
Moderate & 30 & 28 & 29 & 14 & 0 & 20 & 12 & 21 & 6 & 9 & 58 & 42 & 269 \\
Severe & 11 & 50 & 23 & 7 & 0 & 0 & 28 & 6 & 16 & 6 & 0 & 0 & 147 \\
\midrule
Total & 45 & 120 & 66 & 73 & 29 & 51 & 40 & 27 & 27 & 67 & 109 & 42 & 696 \\
\bottomrule
\end{tabular}}
\label{tab:ground_truth_count}
\end{table}

\subsection{Text annotation}
Annotation of BIMCV original medical reports, and readings prepared by experiment participants was performed manually by two experts, in case of disagreement consensus decision was made by a third expert (agreement rate was over 97\%). During text annotation experts were asked to mark whether any condition related to a specific pathology was present (pathology marked as “positive”), or it was clearly stated that there is no description of pathology (marked as “negative”). If pathology or any condition related to this pathology was not mentioned in the report, then the pathology was marked as “not mentioned”. After annotation, the “not mentioned” class was converted to “negative”.

\subsection{DLA system details}
The DLA system is based on two components. First, a set of deep convolutional neural networks based on a Feature Pyramid Network architecture, which are used to automate morphological measurements and/or segment regions of interest on the volumetric CT image. Second, a rule based text-system which generates a text report from the morphological measurements based on international guidelines, see Table \ref{tab:guidelines} for the guidelines, see Table \ref{tab:experiment_form_templates} for example of the generated report.

\begin{table}[ht]
\centering
\caption{Example of the form used in the experiment. Anamnesis was extracted from existing clinical history in both cases.}
\renewcommand{\arraystretch}{1.5}
\begin{tabular}{p{3cm}p{5cm}p{7cm}}
\toprule
Field & "No pathologies present" template & Example report filled from DLA findings \\
\midrule
Medical history (anamnesis) & 73 y.o. man. Complaints of cough and shortness of breath. & 77 y.o. man. Oncology screening. \\
Lungs & No focal or infiltrative changes. The trachea and large bronchi are unchanged. No pleural effusion detected. & A nodule measuring 15x12mm (average size 14mm according to Fleischner), with a volume of 915mm\textsuperscript{3}, is identified in the right lung. Changes in the lungs with a low probability of COVID-19 are determined. Areas of emphysematous changes less than 6\% are identified in the lungs. The percentage of right lung involvement is 2\%, left lung 1\%, both lungs 2\%. The trachea and large bronchi are unchanged. A 17 ml effusion with an average density of 32 HU is identified in the right pleural cavity. No effusion is identified in the left pleural cavity. \\
Mediastinum & Unchanged, no effusion in the pericardial cavity. & Unchanged, no effusion in the pericardial cavity. \\
Cardiovascular System & The aorta and pulmonary trunk are unchanged. & The descending section of the thoracic aorta is expanded to 33 mm (dilation). Other sections of the aorta at the examined level are not expanded. The pulmonary trunk is not expanded (up to 26 mm). Pulmonary-aortic index - 0.722. Coronary artery calcification is determined - Agatston Index: 42, CAC-DRS 1. \\
Lymph Nodes & Not enlarged. & Not enlarged. \\
Soft Tissues & Unchanged. & Unchanged. \\
Skeletal System & No focal, destructive, or traumatic changes. & Consolidated rib fractures: Right: 4 in the anterior third. Left: 2 in the anterior third. Compression deformation of the Th12 vertebrae body is detected - 30.5\%. A decrease in the mineral density of the vertebral bodies' bone tissue is detected: Th11 - 41 HU, Th12 - 42 HU. \\
Abdominal Organs & No changes detected in the scanned area. & No changes detected in the scanned area. \\
Conclusion & No focal or infiltrative changes detected in the lungs. & Changes in the lungs, low probability of COVID-19, have been identified. A nodule in the right lung has been identified. CT in dynamics is recommended in 6 months. Areas of emphysematous changes less than 6\% are identified in the lungs. Right-sided pleural effusion. Dilation of the descending thoracic section of the aorta has been identified. Other sections of the aorta at the examined level are not expanded. Cardiologist consultation is recommended. Agatston Index: 42 (CAC-DRS 1) - minor calcification. Consolidated rib fractures: Right: 4; Left: 2. Compression deformation of the Th12 vertebrae body (Genant 2) has been identified. A decrease in the mineral density of the bone tissue, corresponding to osteoporosis, has been identified. Endocrinologist consultation is recommended. \\
\bottomrule
\end{tabular}
\label{tab:experiment_form_templates}
\end{table}

\begin{figure}
    \includegraphics[width=.4\textwidth]{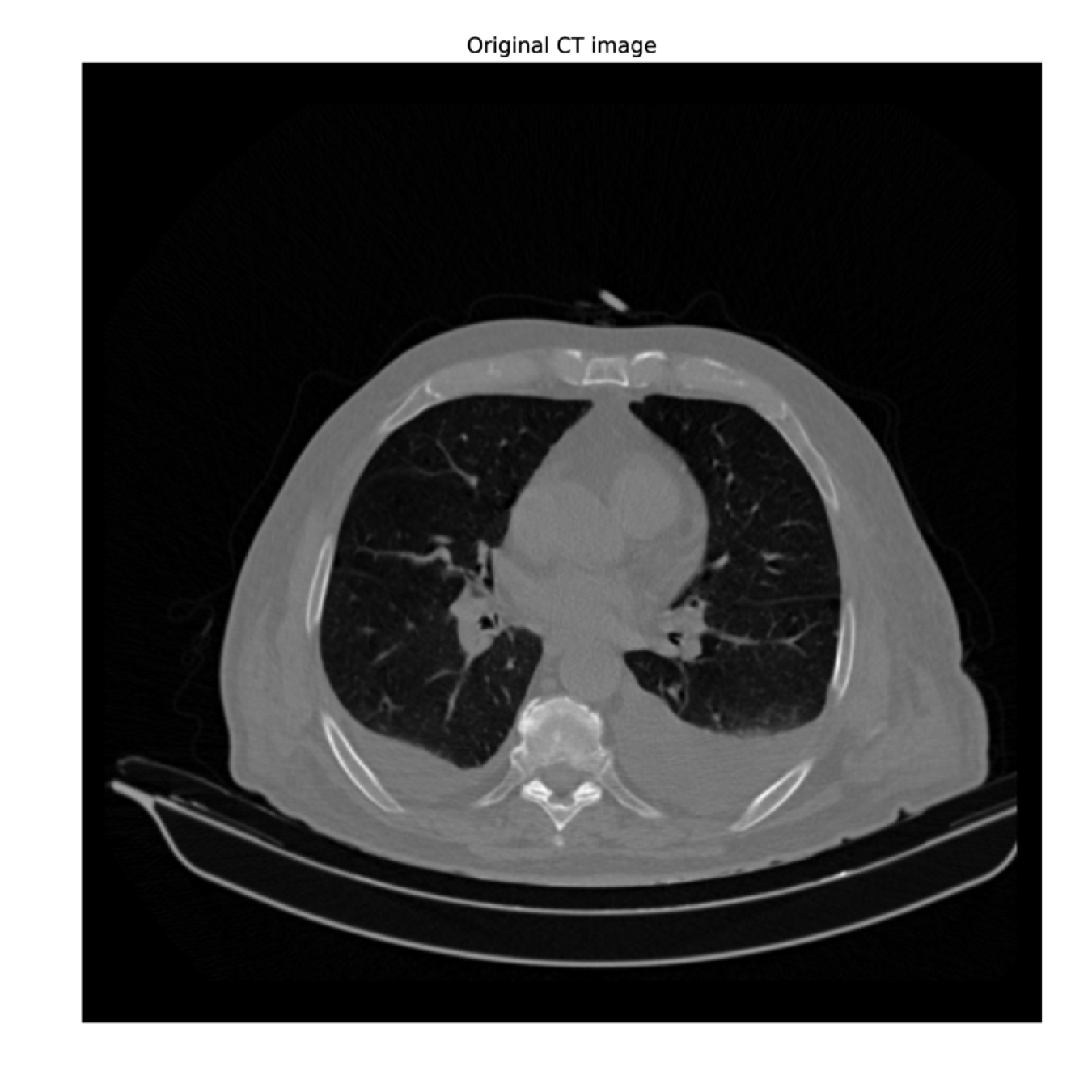}\hfill
    \includegraphics[width=.4\textwidth]{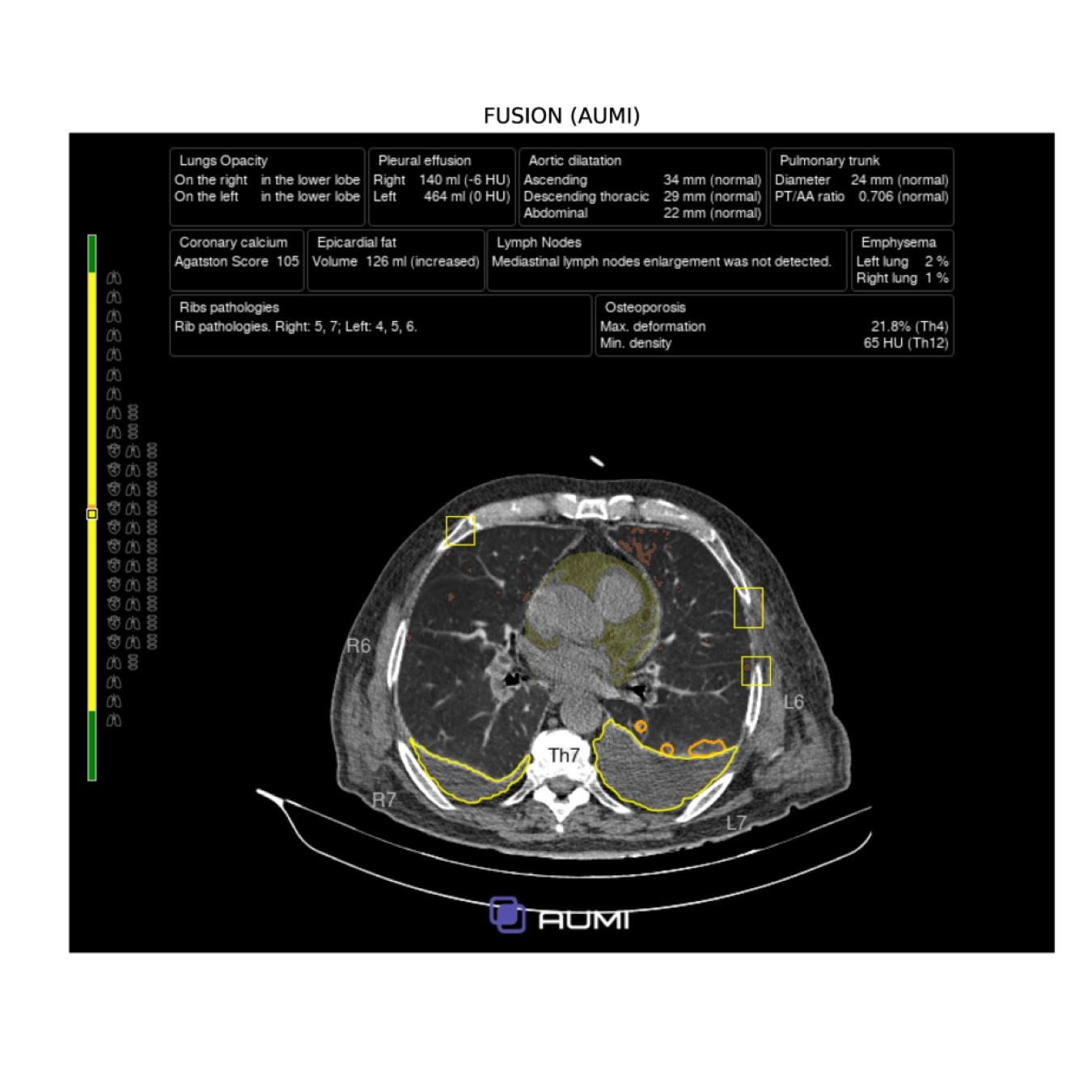}\hfill
    \\[\smallskipamount]
    \includegraphics[width=.4\textwidth]{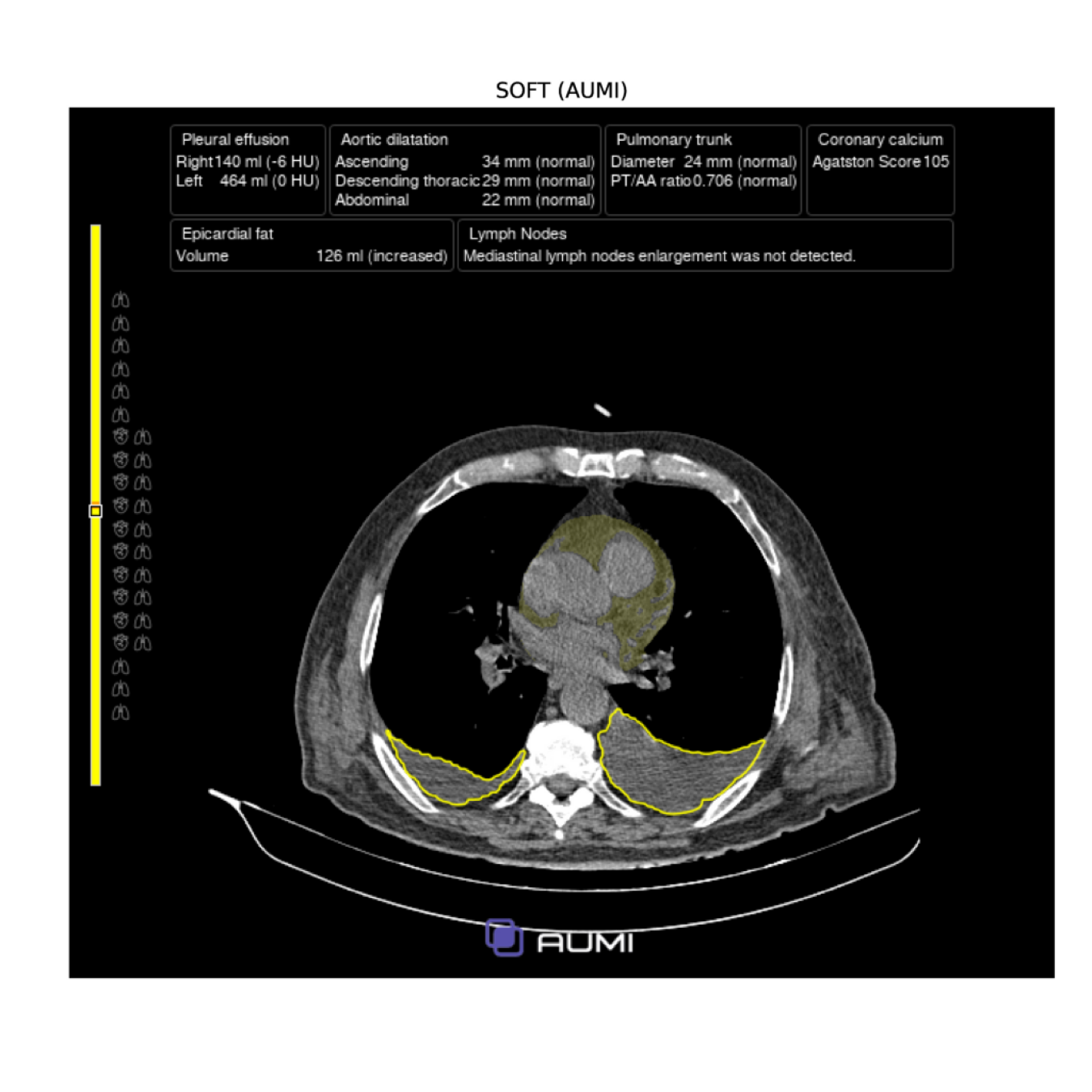}\hfill
    \includegraphics[width=.4\textwidth]{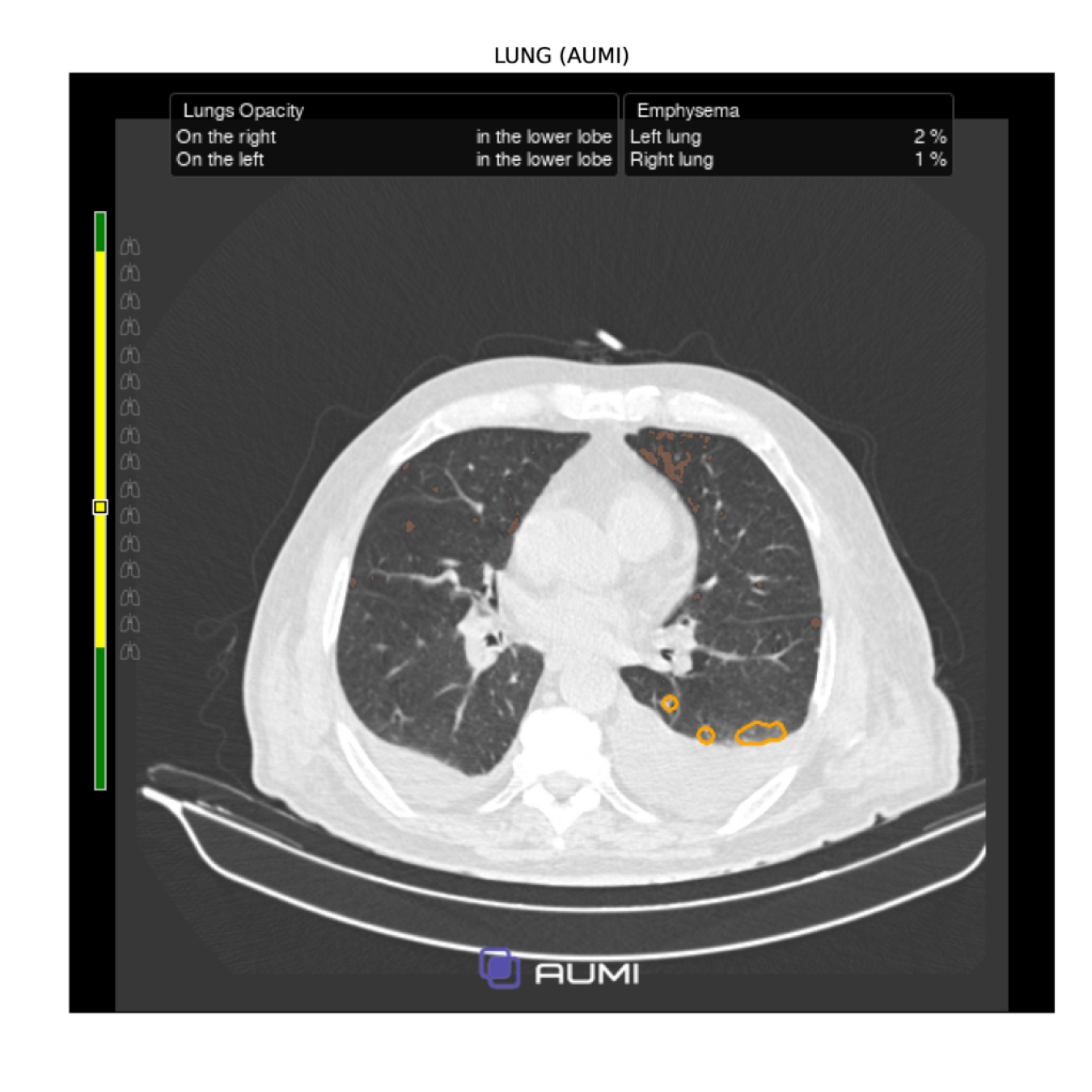}
    \\[\smallskipamount]
    \includegraphics[width=.4\textwidth]{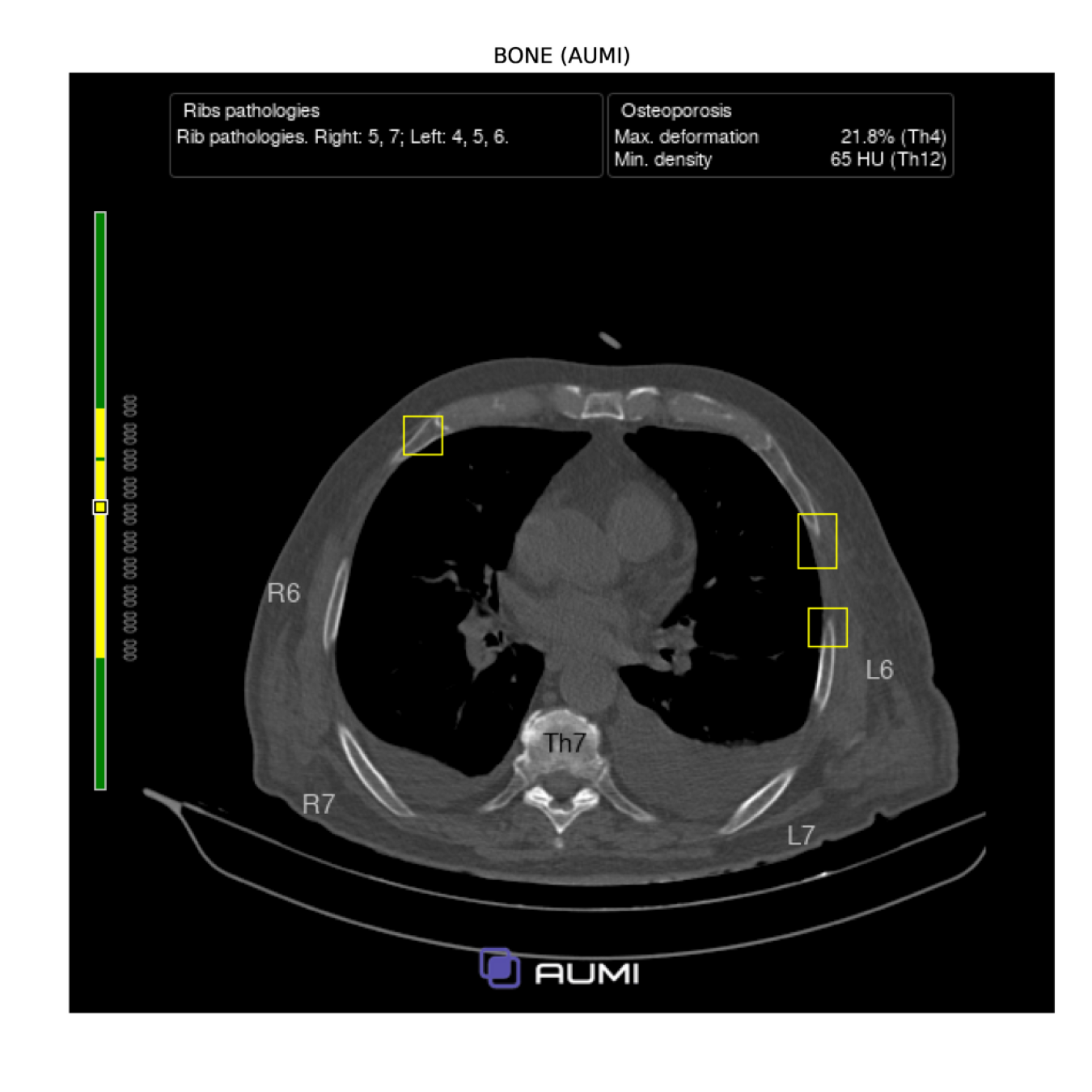}\hfill
    \caption{Original CT image and four CT windows provided by DLA: Fusion, Abdomen (SOFT), Lung, Bone. Best viewed in color.}
    \label{fig:windows}
\end{figure}

\begin{figure}
    \includegraphics[width=.45\textwidth]{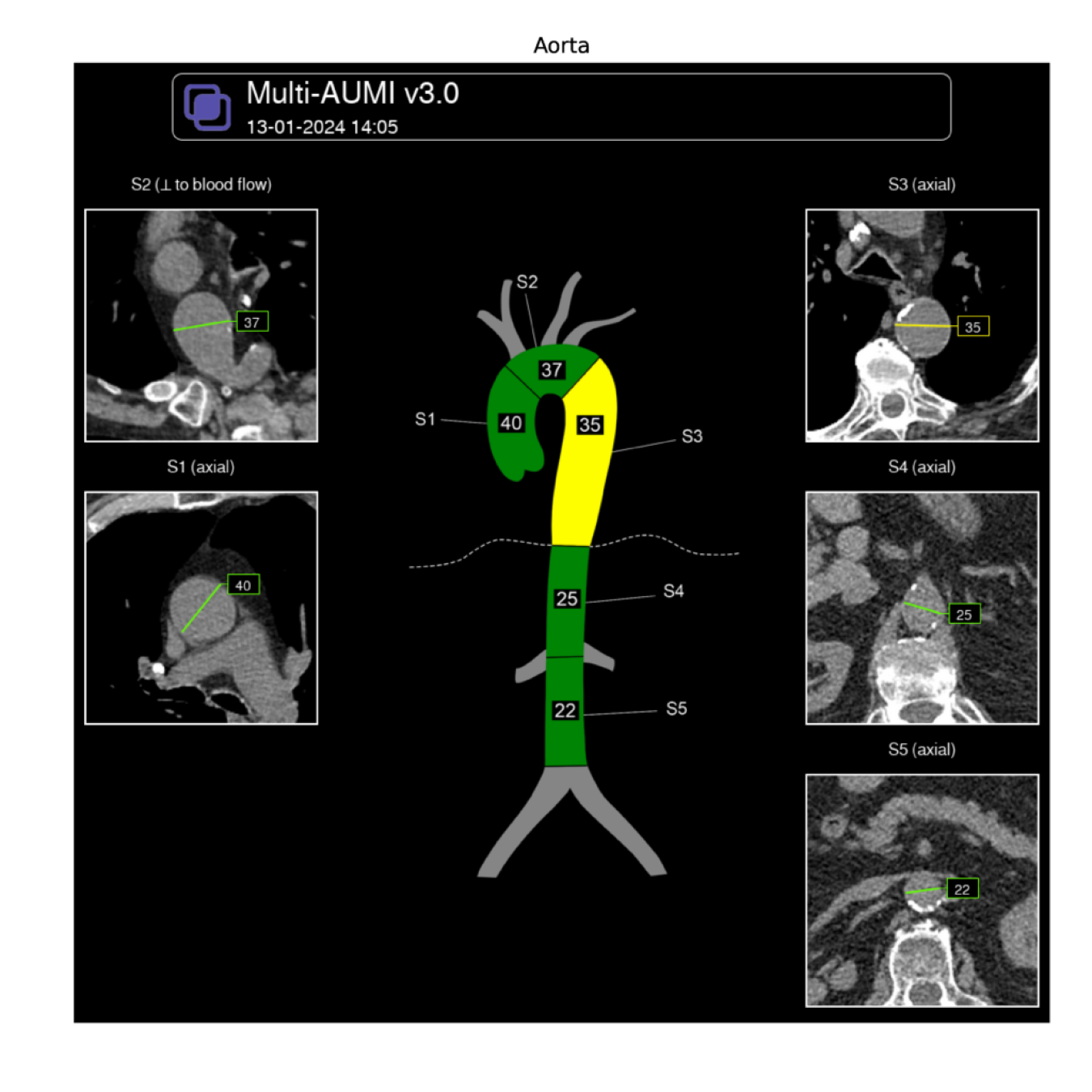}\hfill
    \includegraphics[width=.45\textwidth]{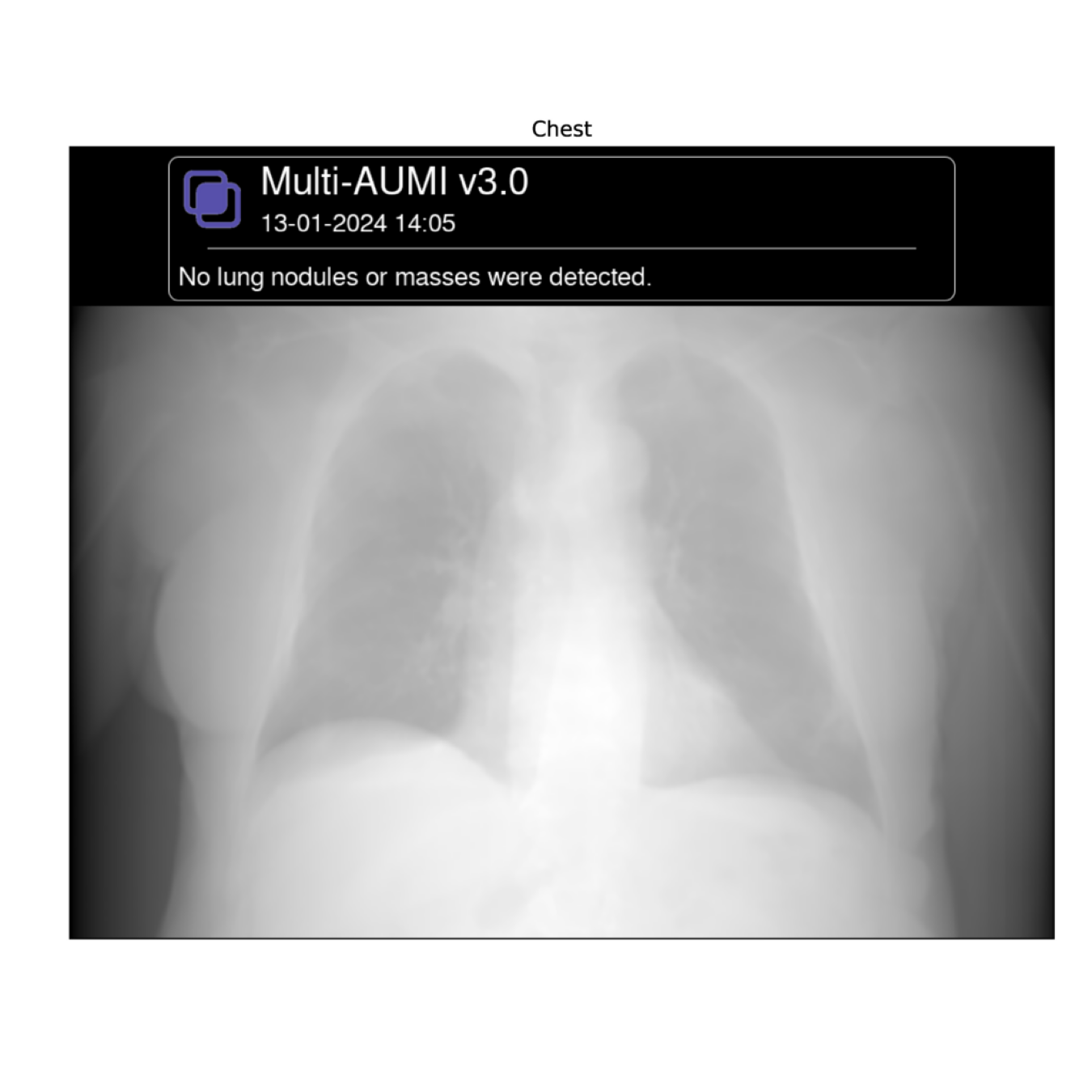}\hfill
    \\[\smallskipamount]
    \includegraphics[width=.45\textwidth]{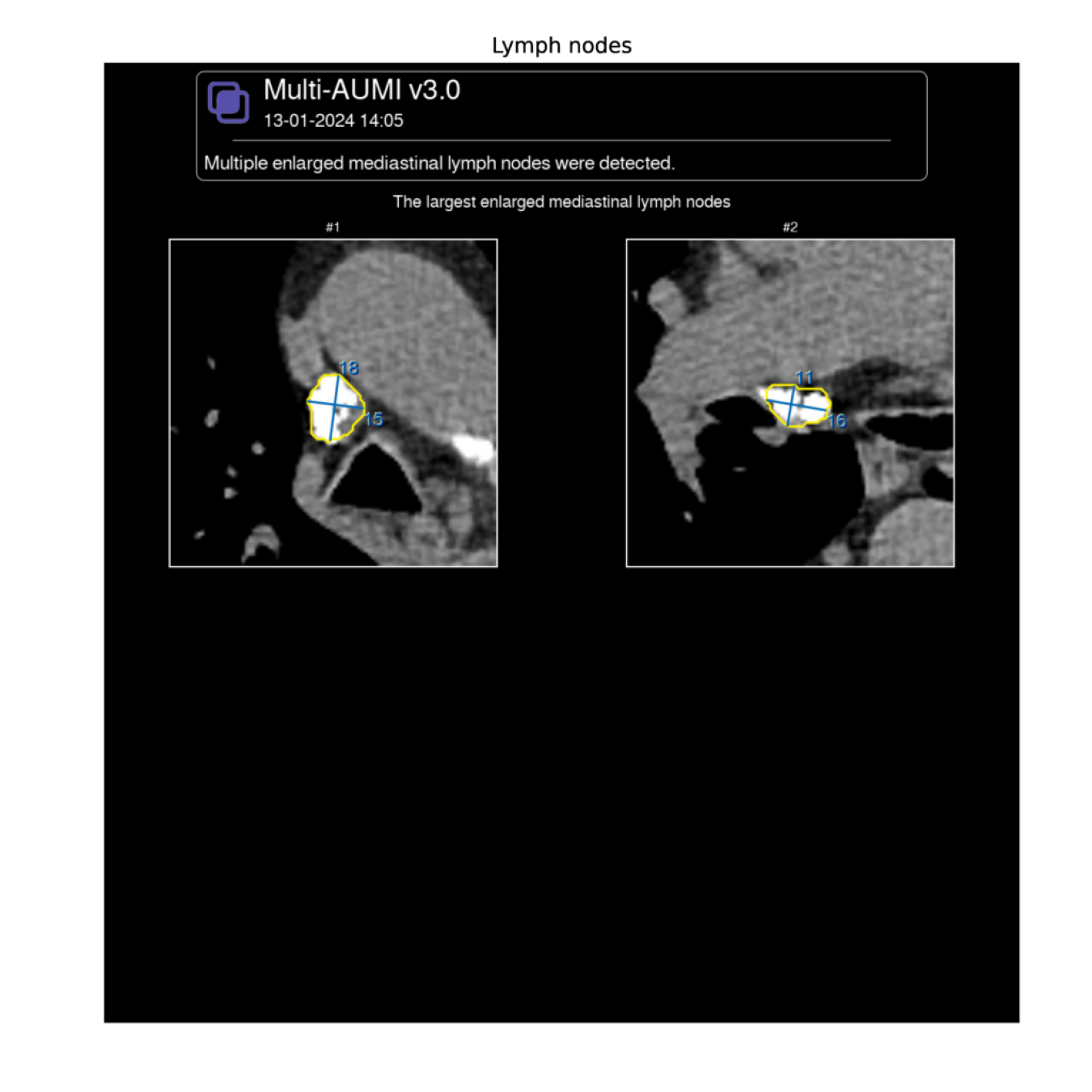}\hfill
    \includegraphics[width=.45\textwidth]{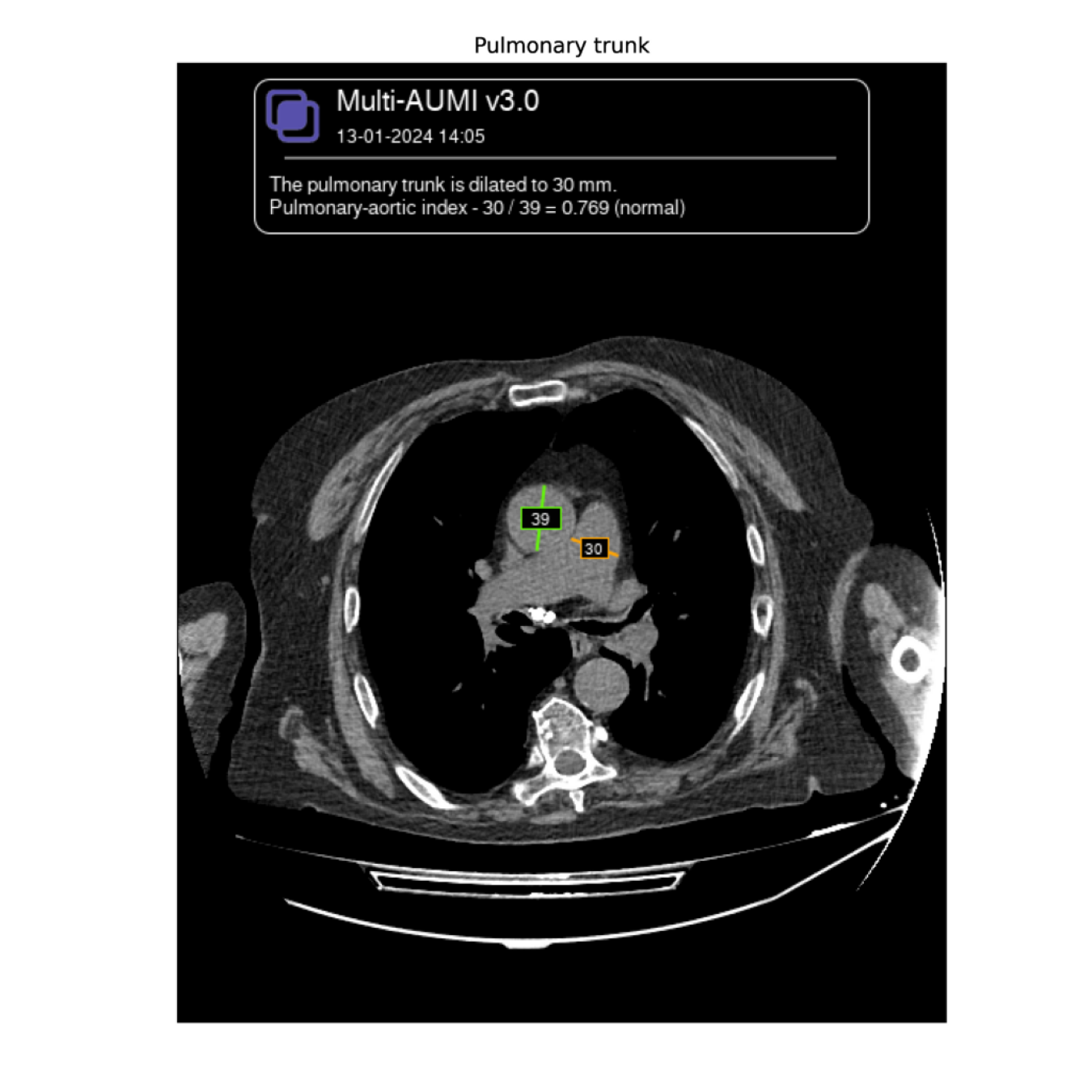}
    \\[\smallskipamount]
    \includegraphics[width=.45\textwidth]{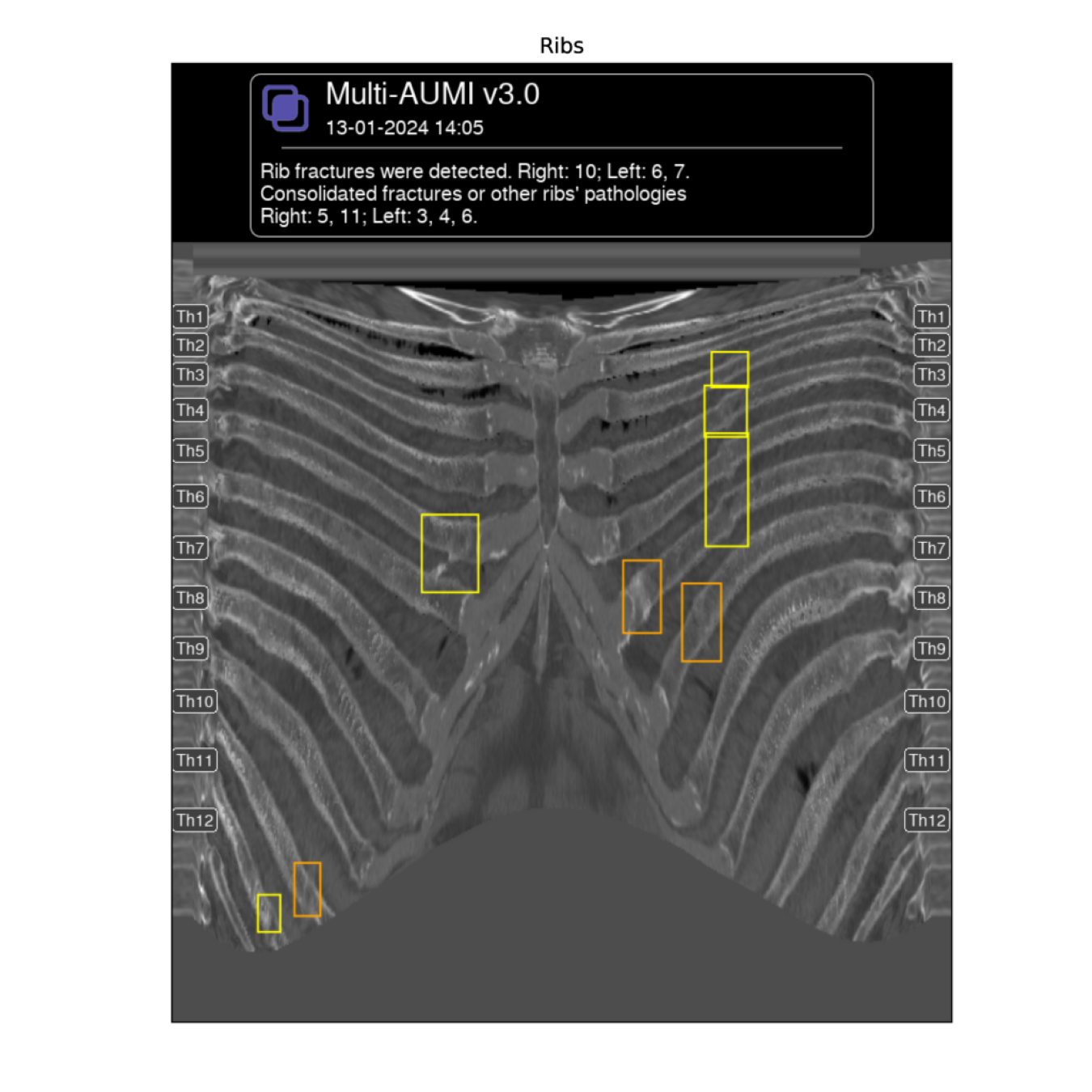}\hfill
    \includegraphics[width=.45\textwidth]{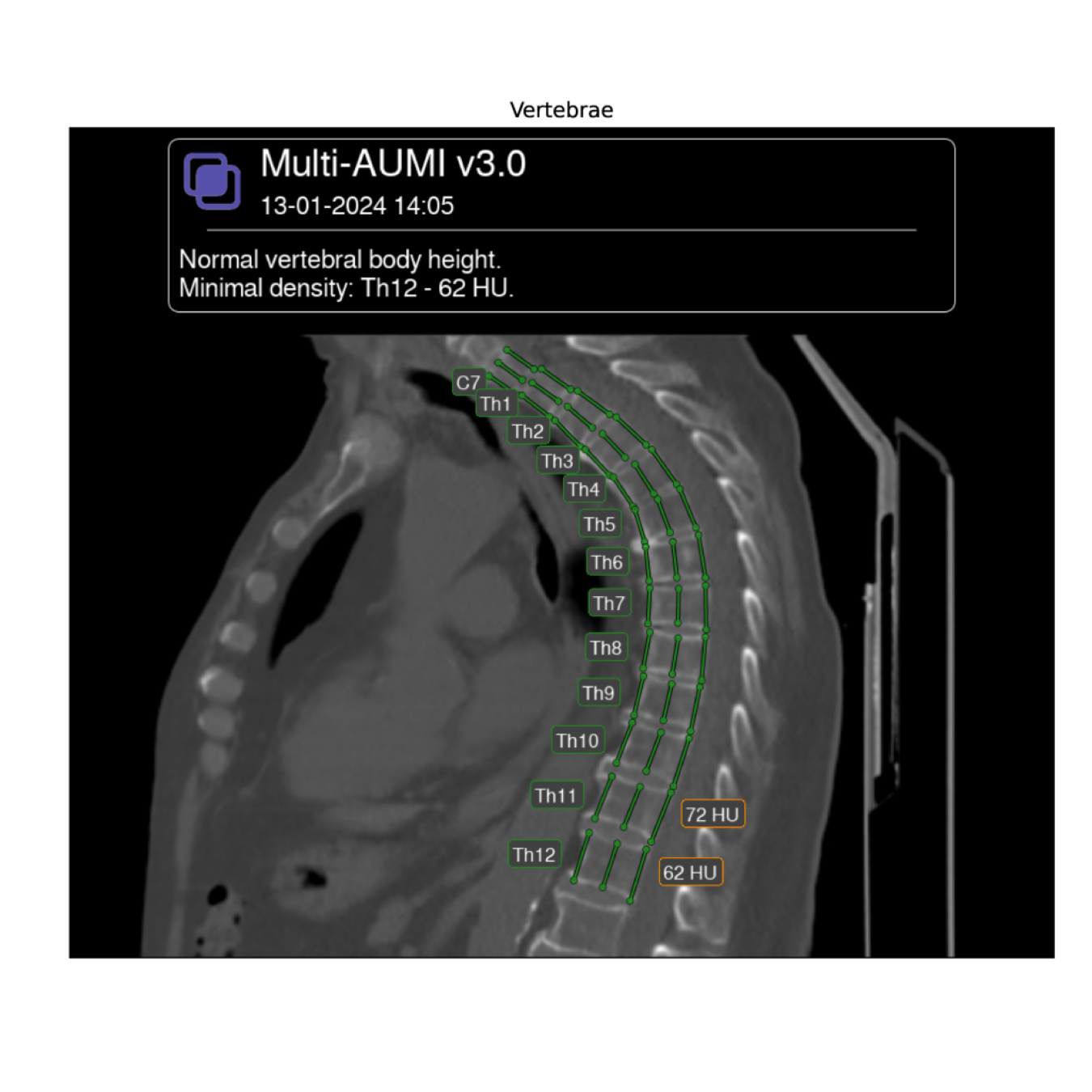}\hfill
    \caption{Summary images provided as first images in processed CT series for each finding: aorta, lungs, mediastinal lymph node, pulmonary trunk, ribs fractures,  vertebrae with Genant index measurements. Best viewed in color.}
    \label{fig:summary}
\end{figure}

\begin{figure}
    \includegraphics[width=.45\textwidth]{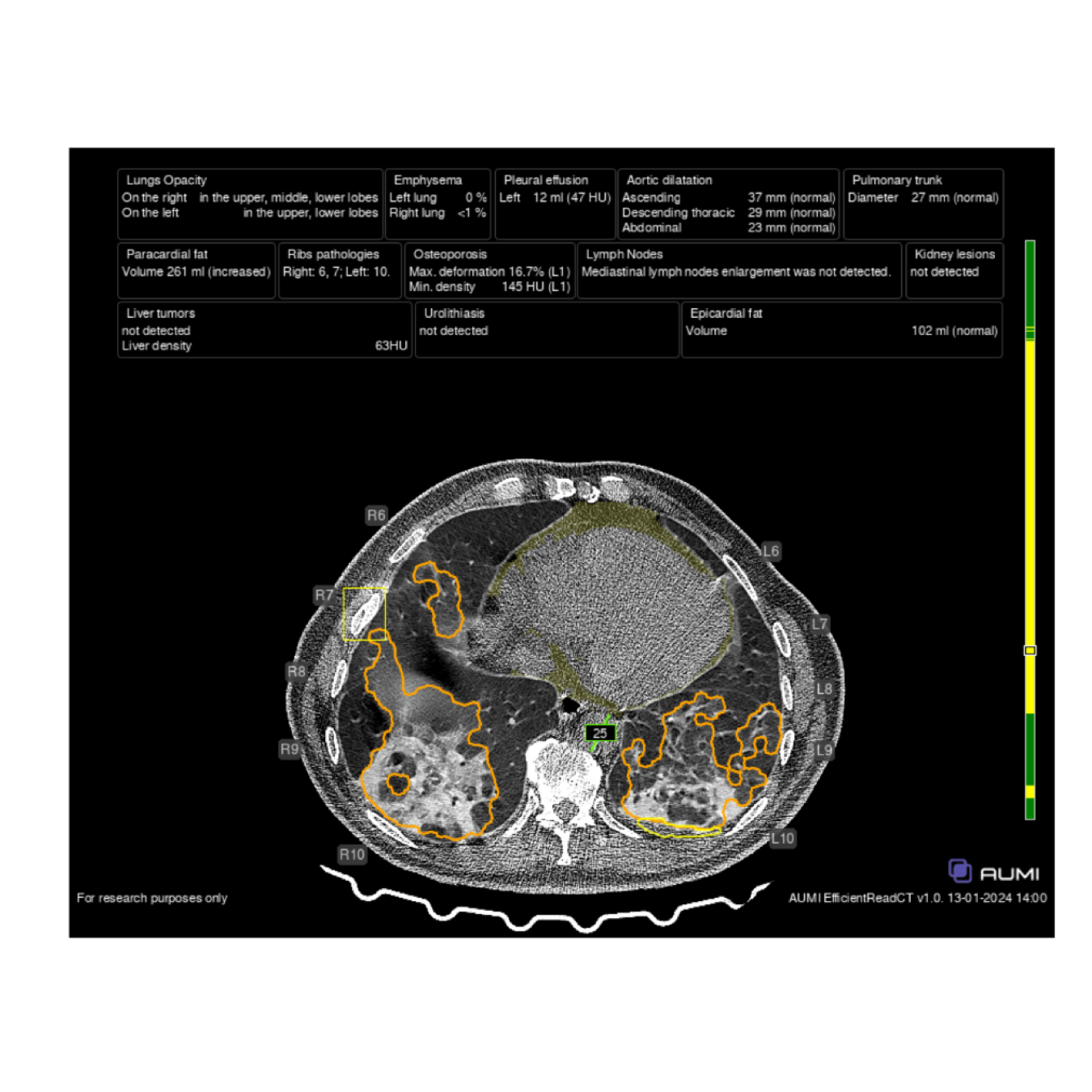}\hfill
    \includegraphics[width=.45\textwidth]{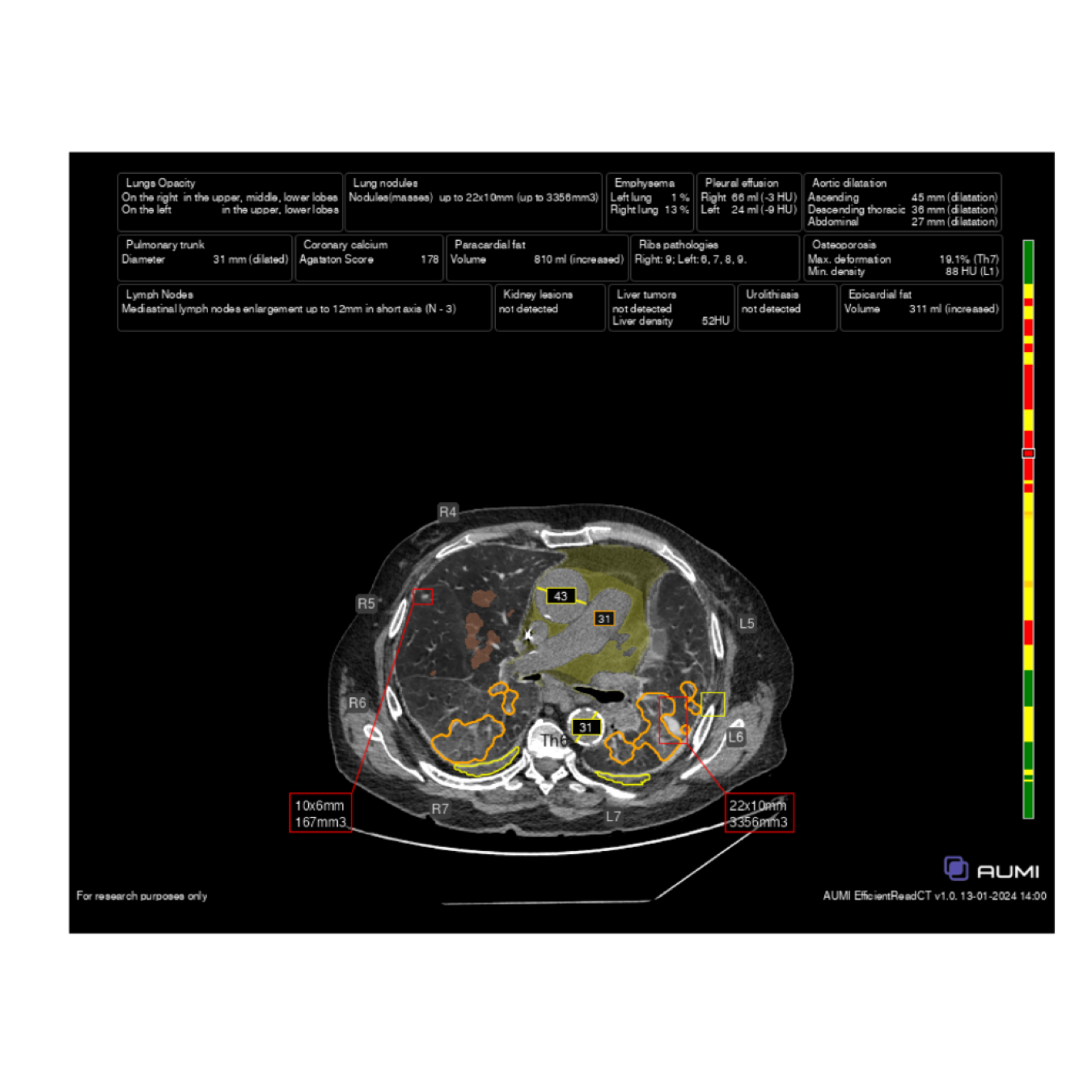}\hfill
    \\[\smallskipamount]
    \includegraphics[width=.45\textwidth]{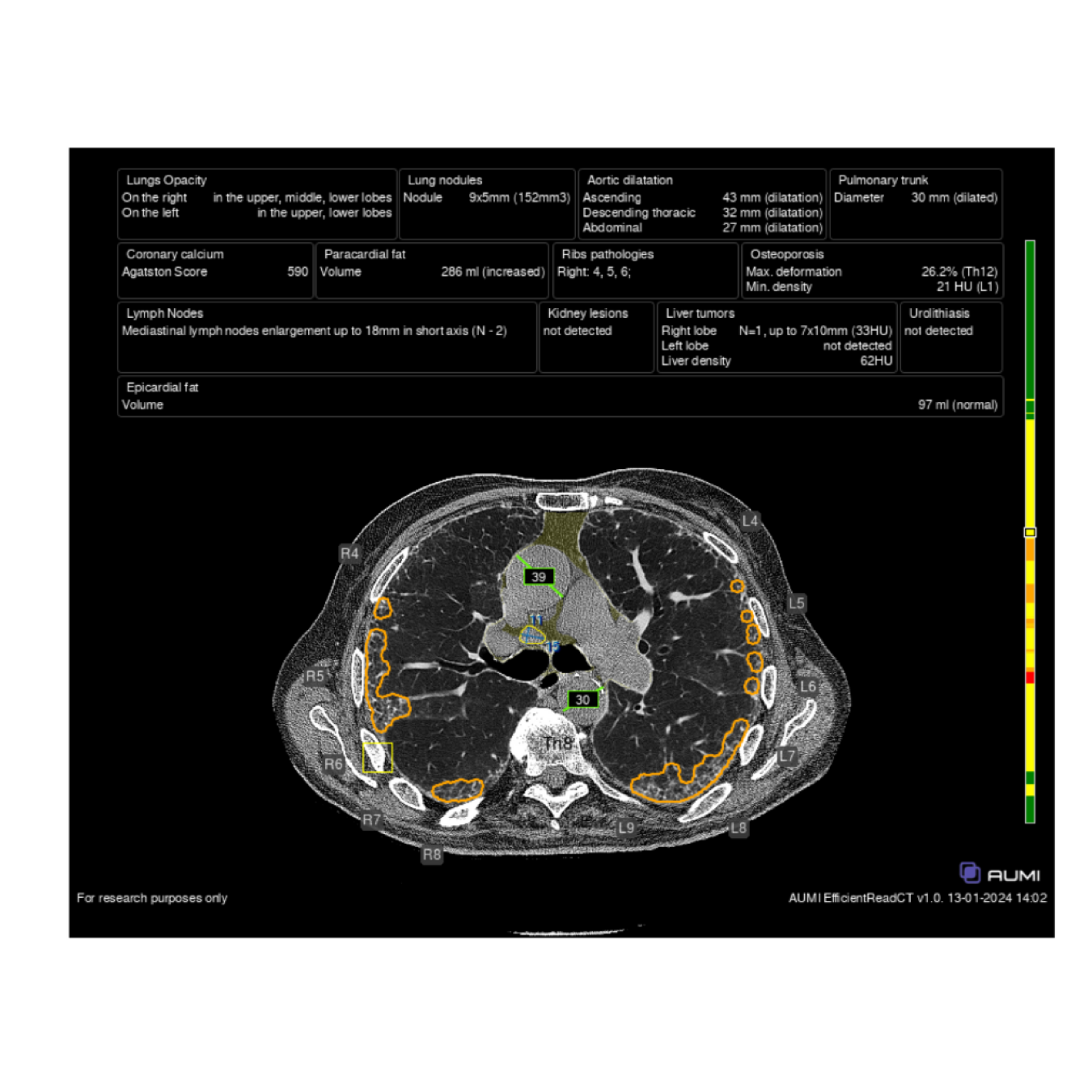}\hfill
    \includegraphics[width=.45\textwidth]{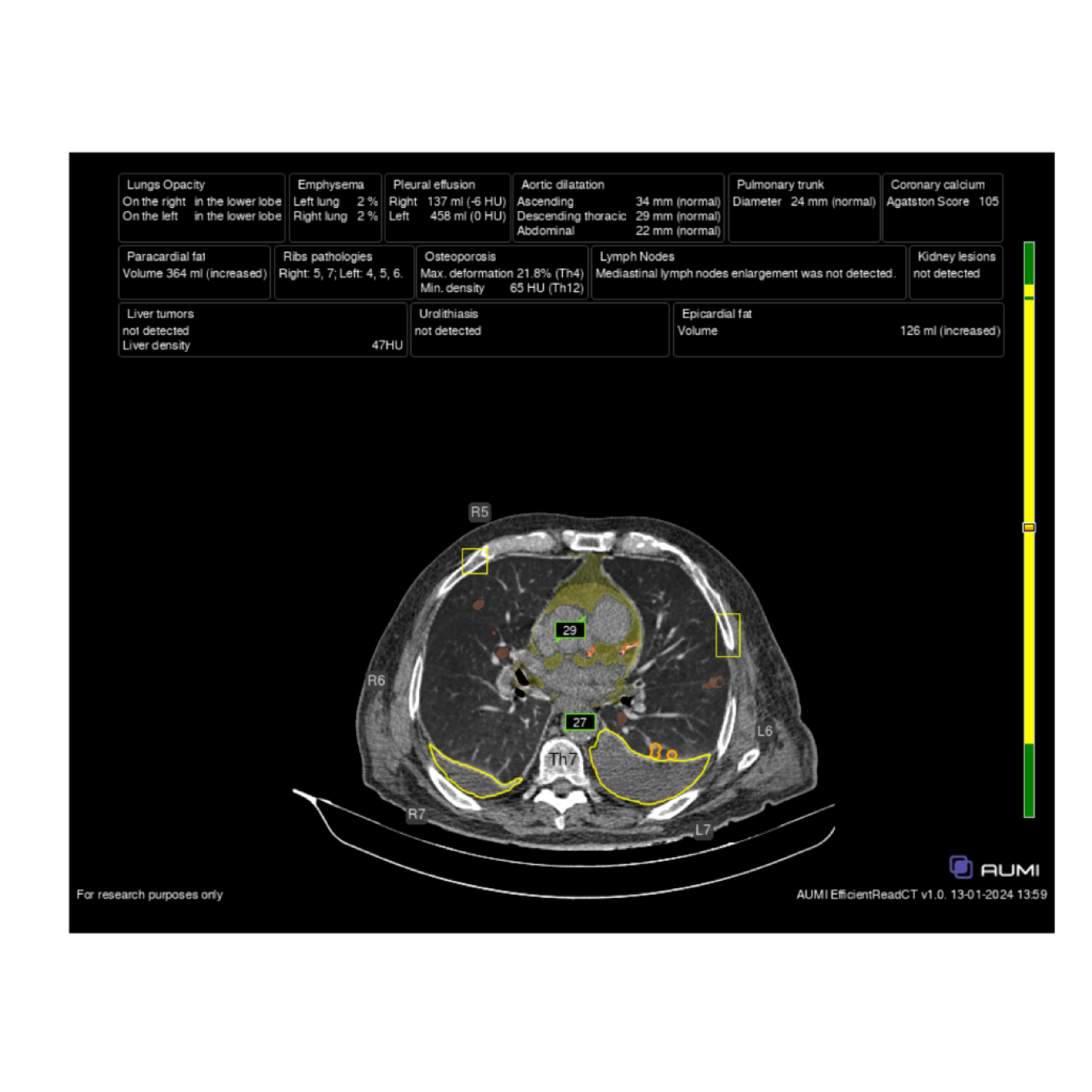}
    \\[\smallskipamount]
    \includegraphics[width=.45\textwidth]{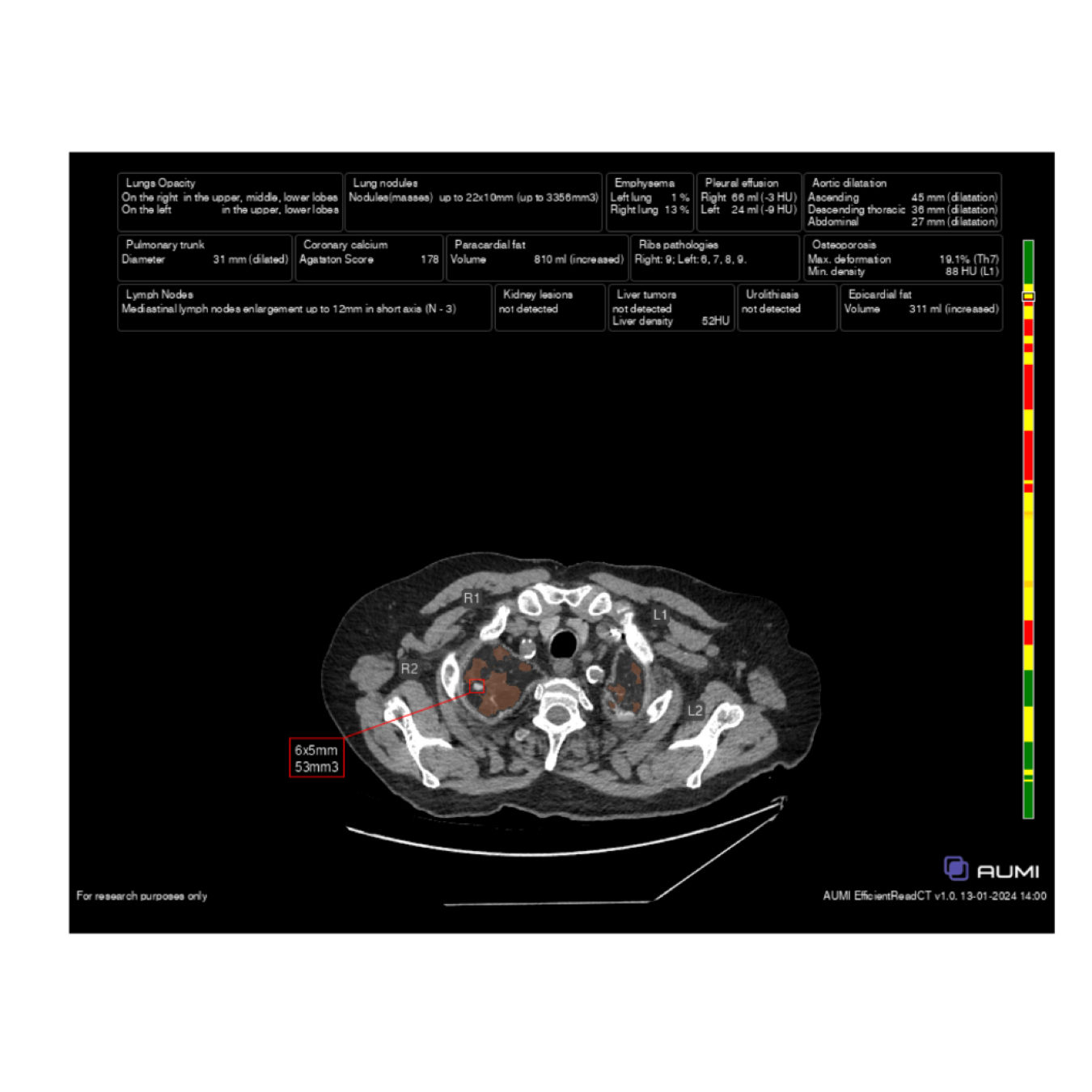}\hfill
    \caption{Example findings. Different DLA findings are overlaid over 2D axial slices of the processed DICOM series. COVID-19 features, consolidated ribs fracture; pulmonary trunk enlargement, aorta enlargement, pleural effusion, lung nodules, features of emphysema, and COVID-19; enlarged lymph nodes, features of COVID-19; pleural effusion, features of emphysema, calcifications; lung nodules, features of emphysema. Best viewed in color.
}
    \label{fig:findings}
\end{figure}

Results of the first component are displayed on a DICOM study in the form of burned-in coloured overlay. Interface highlight schema varies depending on pathology (diameters, contours, or boxes), see Figures \ref{fig:windows}, \ref{fig:summary}, \ref{fig:findings}. A generated text report is provided in the DICOM Structured Report format.

\subsection{BIMCV dataset statistic}
\begin{figure}
    \includegraphics[width=1\textwidth]{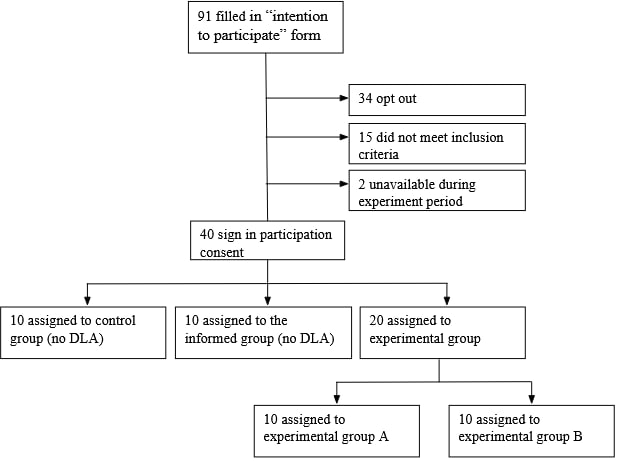}
    \caption{Participated radiologists profile.}
    \label{fig:participants-profile}
\end{figure}

We have used 200 studies from the first release of BIMCV-COVID19 dataset, see Figure \ref{fig:participants-profile} for data selection profile. Table \ref{tab:total_findings_by_pathology} provides information on pathologies frequency (ground truth, and in different experiment arms). Table \ref{tab:severity} provides information of ground truth findings stratified by different severity levels.

\subsection{Participants-wise and pathologies-wise metrics.}
Tables \ref{tab:radiologists_wise_metrics} and \ref{tab:pathology_wise} provide classification metrics for individual radiologists (averaged over 12 pathologies) and for individual pathologies, averaged over radiologists.

\begin{longtable}{llccccc} 
\caption{Pathology-wise metrics in all groups.}\\
\toprule
Group & Pathology & Precision & Sensitivity & Specificity & F1 & Accuracy \\
\midrule
Control (no AI) & adrenal & 83.3 & 12.5 & 99.4 & 21.7 & 82.0 \\
Control (no AI) & aorta & 33.3 & 7.1 & 96.2 & 11.8 & 77.5 \\
Control (no AI) & coronary calcium & 95.9 & 58.3 & 96.2 & 72.5 & 73.5 \\
Control (no AI) & covid & 69.0 & 94.5 & 75.6 & 79.8 & 82.5 \\
Control (no AI) & emphysema & 93.8 & 67.2 & 97.7 & 78.3 & 87.5 \\
Control (no AI) & genant & 88.9 & 29.6 & 99.4 & 44.4 & 90.0 \\
Control (no AI) & lung nodules & 37.5 & 55.6 & 85.5 & 44.8 & 81.5 \\
Control (no AI) & lymph nodes & 88.9 & 53.3 & 98.1 & 66.7 & 88.0 \\
Control (no AI) & pleural effusion & 88.9 & 78.4 & 96.6 & 83.3 & 92.0 \\
Control (no AI) & pulmonary trunk & 94.1 & 24.2 & 99.3 & 38.6 & 74.5 \\
Control (no AI) & ribs & 62.5 & 17.2 & 98.2 & 27.0 & 86.5 \\
Control (no AI) & osteoporosis & - & 0.0 & 100.0 & - & 45.5 \\
With AI-aid training & adrenal & 57.9 & 27.5 & 95.0 & 37.3 & 81.5 \\
With AI-aid training & aorta & 54.8 & 40.5 & 91.1 & 46.6 & 80.5 \\
With AI-aid training & coronary calcium & 93.0 & 55.0 & 93.8 & 69.1 & 70.5 \\
With AI-aid training & covid & 59.3 & 91.8 & 63.8 & 72.0 & 74.0 \\
With AI-aid training & emphysema & 86.2 & 74.6 & 94.0 & 80.0 & 87.5 \\
With AI-aid training & genant & 77.8 & 51.9 & 97.7 & 62.2 & 91.5 \\
With AI-aid training & lung nodules & 51.3 & 74.1 & 89.0 & 60.6 & 87.0 \\
With AI-aid training & lymph nodes & 58.5 & 53.3 & 89.0 & 55.8 & 81.0 \\
With AI-aid training & pleural effusion & 80.9 & 74.5 & 94.0 & 77.6 & 89.0 \\
With AI-aid training & pulmonary trunk & 75.6 & 51.5 & 91.8 & 61.3 & 78.5 \\
With AI-aid training & ribs & 83.3 & 34.5 & 98.8 & 48.8 & 89.5 \\
With AI-aid training & osteoporosis & 100.0 & 12.8 & 100.0 & 22.8 & 52.5 \\
Experimental without AI & adrenal & 79.2 & 47.5 & 96.9 & 59.4 & 87.0 \\
Experimental without AI & aorta & 66.7 & 38.1 & 94.9 & 48.5 & 83.0 \\
Experimental without AI & coronary calcium & 94.5 & 71.7 & 93.8 & 81.5 & 80.5 \\
Experimental without AI & covid & 63.1 & 95.9 & 67.7 & 76.1 & 78.0 \\
Experimental without AI & emphysema & 79.7 & 70.1 & 91.0 & 74.6 & 84.0 \\
Experimental without AI & genant & 78.3 & 66.7 & 97.1 & 72.0 & 93.0 \\
Experimental without AI & lung nodules & 47.4 & 66.7 & 88.4 & 55.4 & 85.5 \\
Experimental without AI & lymph nodes & 69.4 & 55.6 & 92.9 & 61.7 & 84.5 \\
Experimental without AI & pleural effusion & 89.6 & 84.3 & 96.6 & 86.9 & 93.5 \\
Experimental without AI & pulmonary trunk & 81.4 & 53.0 & 94.0 & 64.2 & 80.5 \\
Experimental without AI & ribs & 81.8 & 62.1 & 97.7 & 70.6 & 92.5 \\
Experimental without AI & osteoporosis & 94.4 & 46.8 & 96.7 & 62.6 & 69.5 \\
Experimental with AI & adrenal & 80.0 & 80.0 & 95.0 & 80.0 & 92.0 \\
Experimental with AI & aorta & 80.8 & 100.0 & 93.7 & 89.4 & 95.0 \\
Experimental with AI & coronary calcium & 96.6 & 93.3 & 95.0 & 94.9 & 94.0 \\
Experimental with AI & covid & 68.9 & 97.3 & 74.8 & 80.7 & 83.0 \\
Experimental with AI & emphysema & 86.2 & 83.6 & 93.2 & 84.8 & 90.0 \\
Experimental with AI & genant & 71.1 & 100.0 & 93.6 & 83.0 & 94.5 \\
Experimental with AI & lung nodules & 52.2 & 88.9 & 87.3 & 65.8 & 87.5 \\
Experimental with AI & lymph nodes & 67.9 & 80.0 & 89.0 & 73.5 & 87.0 \\
Experimental with AI & pleural effusion & 88.7 & 92.2 & 96.0 & 90.4 & 95.0 \\
Experimental with AI & pulmonary trunk & 83.3 & 90.9 & 91.0 & 87.0 & 91.0 \\
Experimental with AI & ribs & 51.9 & 96.6 & 84.8 & 67.5 & 86.5 \\
Experimental with AI & osteoporosis & 89.0 & 96.3 & 85.7 & 92.5 & 91.5 \\
Automated (no radiologist) & adrenal & 79.5 & 87.5 & 94.4 & 83.0 & 93.0 \\
Automated (no radiologist) & aorta & 79.2 & 100.0 & 93.0 & 88.4 & 94.5 \\
Automated (no radiologist) & coronary calcium & 98.1 & 84.2 & 97.5 & 90.6 & 89.5 \\
Automated (no radiologist) & covid & 67.0 & 91.8 & 74.0 & 77.5 & 80.5 \\
Automated (no radiologist) & emphysema & 83.3 & 82.1 & 91.7 & 82.7 & 88.5 \\
Automated (no radiologist) & genant & 73.0 & 100.0 & 94.2 & 84.4 & 95.0 \\
Automated (no radiologist) & lung nodules & 54.2 & 96.3 & 87.3 & 69.3 & 88.5 \\
Automated (no radiologist) & lymph nodes & 59.6 & 75.6 & 85.2 & 66.7 & 83.0 \\
Automated (no radiologist) & pleural effusion & 76.9 & 98.0 & 89.9 & 86.2 & 92.0 \\
Automated (no radiologist) & pulmonary trunk & 85.1 & 86.4 & 92.5 & 85.7 & 90.5 \\
Automated (no radiologist) & ribs & 45.9 & 96.6 & 80.7 & 62.2 & 83.0 \\
Automated (no radiologist) & osteoporosis & 74.6 & 80.7 & 67.0 & 77.5 & 74.5 \\
BIMCV annotation & adrenal & 57.1 & 10.0 & 98.1 & 17.0 & 80.5 \\
BIMCV annotation & aorta & 37.5 & 7.1 & 96.8 & 12.0 & 78.0 \\
BIMCV annotation & coronary calcium & 100.0 & 20.0 & 100.0 & 33.3 & 52.0 \\
BIMCV annotation & covid & 73.1 & 93.2 & 80.3 & 81.9 & 85.0 \\
BIMCV annotation & emphysema & 95.7 & 32.8 & 99.2 & 48.9 & 77.0 \\
BIMCV annotation & genant & 75.0 & 11.1 & 99.4 & 19.4 & 87.5 \\
BIMCV annotation & lung nodules & 66.7 & 66.7 & 94.8 & 66.7 & 91.0 \\
BIMCV annotation & lymph nodes & 62.5 & 33.3 & 94.2 & 43.5 & 80.5 \\
BIMCV annotation & pleural effusion & 90.7 & 76.5 & 97.3 & 83.0 & 92.0 \\
BIMCV annotation & pulmonary trunk & 75.0 & 13.6 & 97.8 & 23.1 & 70.0 \\
BIMCV annotation & ribs & 100.0 & 10.3 & 100.0 & 18.8 & 87.0 \\
BIMCV annotation & osteoporosis & 100.0 & 0.9 & 100.0 & 1.8 & 46.0 \\
\bottomrule
\label{tab:pathology_wise}
\end{longtable}

\begin{longtable}{llccccc}
\caption{Radiologists-wise metrics averaged over 12 pathologies.}\\
\toprule
Group & Radiologist & Precision & Sensitivity & Specificity & F1 & Accuracy \\
\midrule
Control (no AI) & 1 & 34.3 & 60.5 & 78.2 & 43.8 & 75.4 \\
Control (no AI) & 2 & 57.7 & 76.9 & 89.1 & 65.9 & 87.1 \\
Control (no AI) & 5 & 30.1 & 78.6 & 75.9 & 43.6 & 76.2 \\
Control (no AI) & 7 & 60.5 & 87.5 & 82.6 & 71.5 & 83.8 \\
Control (no AI) & 8 & 33.8 & 73.3 & 79.5 & 46.3 & 78.8 \\
Control (no AI) & 13 & 52.1 & 78.7 & 82.4 & 62.7 & 81.7 \\
Control (no AI) & 17 & 31.6 & 86.2 & 74.4 & 46.3 & 75.8 \\
Control (no AI) & 25 & 43.6 & 73.9 & 77.3 & 54.8 & 76.7 \\
Control (no AI) & 28 & 51.0 & 83.3 & 88.6 & 63.3 & 87.9 \\
Control (no AI) & 36 & 40.7 & 84.6 & 76.1 & 55.0 & 77.5 \\
With AI-aid training & 0 & 62.8 & 74.2 & 83.3 & 68.1 & 80.8 \\
With AI-aid training & 4 & 46.2 & 81.1 & 82.8 & 58.8 & 82.5 \\
With AI-aid training & 9 & 44.2 & 69.7 & 86.0 & 54.1 & 83.8 \\
With AI-aid training & 11 & 61.2 & 83.3 & 90.7 & 70.6 & 89.6 \\
With AI-aid training & 18 & 65.8 & 73.2 & 84.0 & 69.3 & 80.8 \\
With AI-aid training & 21 & 55.6 & 61.6 & 78.4 & 58.4 & 73.3 \\
With AI-aid training & 22 & 53.5 & 84.4 & 83.1 & 65.5 & 83.3 \\
With AI-aid training & 35 & 49.4 & 81.6 & 78.5 & 61.5 & 79.2 \\
With AI-aid training & 37 & 61.6 & 75.0 & 84.4 & 67.7 & 82.1 \\
With AI-aid training & 39 & 19.4 & 34.2 & 73.3 & 24.8 & 67.1 \\
Experimental without AI & 3 & 81.6 & 88.6 & 91.8 & 84.9 & 90.8 \\
Experimental without AI & 6 & 65.9 & 84.4 & 84.1 & 74.0 & 84.2 \\
Experimental without AI & 10 & 69.2 & 85.7 & 91.9 & 76.6 & 90.8 \\
Experimental without AI & 12 & 62.5 & 75.0 & 91.0 & 68.2 & 88.3 \\
Experimental without AI & 14 & 85.2 & 62.2 & 95.2 & 71.9 & 85.0 \\
Experimental without AI & 15 & 77.5 & 86.1 & 89.3 & 81.6 & 88.3 \\
Experimental without AI & 16 & 78.0 & 86.5 & 89.2 & 82.1 & 88.3 \\
Experimental without AI & 19 & 60.0 & 85.7 & 82.6 & 70.6 & 83.3 \\
Experimental without AI & 20 & 81.8 & 81.8 & 95.9 & 81.8 & 93.3 \\
Experimental without AI & 23 & 41.0 & 69.6 & 76.3 & 51.6 & 75.0 \\
Experimental without AI & 24 & 40.5 & 78.9 & 78.2 & 53.6 & 78.3 \\
Experimental without AI & 26 & 71.8 & 82.4 & 87.2 & 76.7 & 85.8 \\
Experimental without AI & 27 & 57.1 & 76.2 & 87.9 & 65.3 & 85.8 \\
Experimental without AI & 29 & 61.4 & 75.0 & 79.8 & 67.5 & 78.3 \\
Experimental without AI & 30 & 77.8 & 81.4 & 87.0 & 79.5 & 85.0 \\
Experimental without AI & 31 & 85.2 & 71.9 & 95.5 & 78.0 & 89.2 \\
Experimental without AI & 32 & 53.3 & 69.6 & 85.6 & 60.4 & 82.5 \\
Experimental without AI & 33 & 46.4 & 61.9 & 84.8 & 53.1 & 80.8 \\
Experimental without AI & 34 & 51.2 & 78.6 & 77.2 & 62.0 & 77.5 \\
Experimental without AI & 38 & 43.2 & 64.0 & 77.9 & 51.6 & 75.0 \\
Experimental with AI & 3 & 95.3 & 78.8 & 97.1 & 86.3 & 89.2 \\
Experimental with AI & 6 & 85.0 & 91.9 & 92.8 & 88.3 & 92.5 \\
Experimental with AI & 10 & 97.4 & 86.4 & 98.7 & 91.6 & 94.2 \\
Experimental with AI & 12 & 96.4 & 73.0 & 98.8 & 83.1 & 90.8 \\
Experimental with AI & 14 & 97.7 & 81.1 & 98.5 & 88.7 & 90.8 \\
Experimental with AI & 15 & 94.9 & 86.0 & 97.4 & 90.2 & 93.3 \\
Experimental with AI & 16 & 100.0 & 82.2 & 100.0 & 90.2 & 93.3 \\
Experimental with AI & 19 & 97.6 & 88.9 & 98.7 & 93.0 & 95.0 \\
Experimental with AI & 20 & 96.3 & 74.3 & 98.8 & 83.9 & 91.7 \\
Experimental with AI & 23 & 90.0 & 76.6 & 94.5 & 82.8 & 87.5 \\
Experimental with AI & 24 & 87.8 & 76.6 & 93.2 & 81.8 & 86.7 \\
Experimental with AI & 26 & 88.5 & 69.7 & 96.6 & 78.0 & 89.2 \\
Experimental with AI & 27 & 83.3 & 71.4 & 95.7 & 76.9 & 90.0 \\
Experimental with AI & 29 & 85.2 & 74.2 & 95.5 & 79.3 & 90.0 \\
Experimental with AI & 30 & 100.0 & 77.8 & 100.0 & 87.5 & 93.3 \\
Experimental with AI & 31 & 100.0 & 55.0 & 100.0 & 71.0 & 85.0 \\
Experimental with AI & 32 & 89.2 & 78.6 & 94.9 & 83.5 & 89.2 \\
Experimental with AI & 33 & 82.2 & 86.0 & 89.6 & 84.1 & 88.3 \\
Experimental with AI & 34 & 86.8 & 89.2 & 94.0 & 88.0 & 92.5 \\
Experimental with AI & 38 & 86.7 & 74.3 & 95.3 & 80.0 & 89.2 \\
\bottomrule
\label{tab:radiologists_wise_metrics}
\end{longtable}

\end{document}